\documentclass[aps,prd,preprint,preprintnumbers,nofootinbib,superscriptaddress]{revtex4}
\usepackage{ae}
\usepackage{bm} 
\usepackage{color}
\usepackage{amsmath}
\usepackage{amssymb}
\usepackage{amsfonts}
\usepackage{graphicx}
\usepackage{slashed}
\usepackage{setspace}
\usepackage{stackrel}

\newcommand{\Eqref}[1]{Eq.~\eqref{#1}}
\newcommand{\Tr}{\text{Tr}} 

\newcommand{\F}{{\cal F}}
\newcommand{\G}{{\cal G}}
\newcommand{\dLF}{\frac{\partial{\cal L}}{\partial{\cal F}}}
\newcommand{\dLG}{\frac{\partial{\cal L}}{\partial{\cal G}}}
\newcommand{\dLFF}{\frac{\partial^2{\cal L}}{\partial{\cal F}^2}}
\newcommand{\dLGG}{\frac{\partial^2{\cal L}}{\partial{\cal G}^2}}
\newcommand{\dLFG}{\frac{\partial^2{\cal L}}{\partial{\cal F}\partial{\cal G}}}
\newcommand{\dLFFF}{\frac{\partial^3{\cal L}}{\partial{\cal F}^3}}
\newcommand{\dLGGG}{\frac{\partial^3{\cal L}}{\partial{\cal G}^3}}
\newcommand{\dLFFG}{\frac{\partial^3{\cal L}}{\partial{\cal F}^2\partial{\cal G}}}
\newcommand{\dLFGG}{\frac{\partial^{3}{\cal L}}{\partial{\cal F}\partial{\cal G}^{2}}}
\newcommand{\Abar}{\bar{A}}
\newcommand{\Fbar}{\bar{F}}
\newcommand{\Jbar}{\bar{J}}

\allowdisplaybreaks

\usepackage{ulem}
\usepackage{color}

\begin{document}

\setlength{\unitlength}{1mm}
\title{An Addendum to the Heisenberg-Euler Effective Action Beyond One Loop}
\author{Holger Gies}\email{holger.gies@uni-jena.de}
\author{Felix Karbstein}\email{felix.karbstein@uni-jena.de}
\affiliation{Helmholtz-Institut Jena, Fr\"obelstieg 3, 07743 Jena, Germany}
\affiliation{Theoretisch-Physikalisches Institut, Abbe Center of Photonics, \\ Friedrich-Schiller-Universit\"at Jena, Max-Wien-Platz 1, 07743 Jena, Germany}

\date{\today}

\begin{abstract}
 We study the effective interactions of external electromagnetic fields induced by fluctuations of virtual particles in the vacuum of quantum electrodynamics.
 Our main focus is on these interactions at two-loop order.
 We discuss in detail the emergence of the renowned Heisenberg-Euler effective action from the underlying microscopic theory of quantum electrodynamics,
 emphasizing its distinction from a standard one-particle irreducible effective action.
 In our explicit calculations we limit ourselves to constant and slowly varying external fields, allowing us to adopt a locally constant field approximation.
 One of our main findings is that at two-loop order there is a finite one-particle reducible contribution to the Heisenberg-Euler effective action in constant fields, which was previously assumed to vanish.
 In addition to their conceptual significance, our results are relevant for high-precision probes of quantum vacuum nonlinearity in strong electromagnetic fields.
\end{abstract}

\maketitle

%%%%%%%%%%%%%%%%%%%%%%%%%%% 
\section{Introduction} \label{sec:intro}
%%%%%%%%%%%%%%%%%%%%%%%%%%%

One of the striking predictions of quantum field theory (QFT) is that virtual charged particle-antiparticle fluctuations in the quantum vacuum can induce nonlinear interactions among electromagnetic fields
\cite{Euler:1935zz,Heisenberg:1935qt,Weisskopf}; for reviews emphasizing various theoretical aspects as well as prospects for the experimental detection of such effects, see \cite{Dittrich:1985yb,Dittrich:2000zu,Marklund:2008gj,Dunne:2008kc,Heinzl:2008an,DiPiazza:2011tq,Dunne:2012vv,Battesti:2012hf,King:2015tba,Karbstein:2016hlj}.
Aiming at probing the vacuum of the Standard Model of particle physics with classical electromagnetic fields and low energy photons, the dominant effective interactions are governed by quantum electrodynamics (QED).

For the macroscopic electromagnetic fields presently attainable in the laboratory, the effects of QED vacuum nonlinearities are rather small, making their experimental detection challenging \cite{Battesti:2012hf,DiPiazza:2011tq}.
These effective interactions have no tree-level analogue, but are mediated by at least one electron-positron loop.
For electromagnetic fields which vary on scales much larger than the Compton wavelength of the electron $\lambdabar_{\rm C}=3.86 \cdot 10^{-13}{\rm m}$ and are weak in comparison to the critical electric field strength $E_{\rm cr}\equiv\frac{m^2c^3}{e\hbar}\simeq1.3\cdot10^{18}{\rm V}/{\rm m}$ \cite{Sauter:1931zz,Heisenberg:1935qt,Schwinger:1951nm}, i.e., fulfill $\{|\vec{E}|,c|\vec{B}|\}\ll E_{\rm cr}$,
this results in a parametric suppression of the effective interaction by inverse powers of the electron mass. 
Most of the electromagnetic fields available in the laboratory meet this criterion.

The effective interactions can in particular impact photon propagation and give rise to remarkable effects such as vacuum birefringence experienced by probe photons traversing a classical electromagnetic field \cite{Toll:1952,Baier,BialynickaBirula:1970vy};
for ongoing experimental efforts aiming at the verification of vacuum birefringence using macroscopic fields, see \cite{Cantatore:2008zz,Berceau:2011zz}; for theoretical proposals advocating dedicated high-intensity laser experiments, see \cite{Kotkin:1996nf,Heinzl:2006xc,DiPiazza:2006pr,Dinu:2013gaa,Ilderton:2016khs,King:2016jnl,Schlenvoigt:2016,Karbstein:2015xra,Karbstein:2016lby}.
Recently, indications have been reported for the relevance of QED vacuum birefringence for optical polarimetry of a neutron star \cite{Mignani:2016fwz}.
Other theoretical proposals for optical signatures of quantum vacuum nonlinearity have focused on photon-photon scattering in the form of laser-pulse collisions \cite{Lundstrom:2005za,Lundin:2006wu,Tommasini:2010fb,King:2012aw}, interference effects \cite{King:2013am,Hatsagortsyan:2011,Sarazin:2016zer}, quantum reflection \cite{Gies:2013yxa}, photon merging \cite{Yakovlev:1966,DiPiazza:2007cu,Gies:2014jia,Gies:2016czm},
photon splitting \cite{BialynickaBirula:1970vy,Adler:1971wn,Adler:1970gg,Papanyan:1971cv,Stoneham:1979,Baier:1986cv,Adler:1996cja,DiPiazza:2007yx}, and higher-harmonic generation from laser driven vacuum \cite{DiPiazza:2005jc,Fedotov:2006ii,Karbstein:2014fva,Bohl:2015uba}.
Finally, and perhaps most strikingly, strong electric fields can facilitate the spontaneous formation of real electron-position pairs from the QED vacuum via the Schwinger effect \cite{Sauter:1931zz,Heisenberg:1935qt,Schwinger:1951nm}.

It is a fascinating aspect of this plethora of phenomena that they manifest the effective interactions of electromagnetic fields beyond Maxwell's linear theory, which can be summarized elegantly by an effective action that dates back to the early days of quantum field theory \cite{Heisenberg:1935qt,Weisskopf}: the Heisenberg-Euler effective action.
Its matured embedding into the modern language of field theory is due to Schwinger \cite{Schwinger:1951nm}, who gave a nonperturbative definition of this action by means of the vacuum persistence amplitude, i.e., the Schwinger functional.
Nowadays, QFT is often defined in terms of generating functionals for correlation functions, with the concept of the effective action being identified with the generating functions of one-particle irreducible (1PI) correlators (proper vertices).

In this article, we emphasize that the Heisenberg-Euler effective action is different from -- though related to -- the 1PI effective action.
This fact has, of course, been well known in the specialized literature \cite{Ritus:1975,Dittrich:1985yb} but is sometimes confused in textbooks.
We detail the construction of the Heisenberg-Euler effective action from the standard definition of QED in terms of the partition function in the present work.
The difference between the two effective actions is manifested by one-particle reducible (1PR) contributions to the Heisenberg-Euler action.
In a perturbative loop expansion, such 1PR contributions occur at and beyond the two-loop order.
At two-loop order, we find that there is a finite 1PR contribution to the Heisenberg-Euler effective action in constant electromagnetic fields, which was previously believed to vanish.

Using a locally constant field approximation (LCFA), we also
study in detail the effective theory of slowly varying classical
background fields and low-frequency photon fields in the QED vacuum.
The photon polarization tensor derived within the latter contains 1PI, 1PR, as well as disconnected
contributions, all of which can be understood as generated by the 1PI effective action.

As higher-loop diagrams are typically suppressed in comparison to the one-loop diagram, a proper inclusion of the previously neglected 1PR diagrams is expected to impact the proposed experimental signatures of quantum vacuum nonlinearities only at subleading order. In fact, most of the theoretical studies listed above exclusively limit themselves to one-loop order. For instance for vacuum birefringence in weak fields, the two-loop contribution represents only a 1\%\ correction \cite{Dittrich:1998fy}.

Our article is organized as follows. Section~\ref{sec:HEvs1PI} is devoted to an in-depth discussion of the Heisenberg-Euler effective action.
Here, we elaborate differences and common ground with respect to the standardized 1PI effective action and show how the Heisenberg-Euler effective action emerges from the microscopic theory of QED.
Finally, we explicitly sketch its diagrammatic expansion up to two-loop order.
In Sec.~\ref{sec:knownHE} we focus on the Heisenberg-Euler effective action in constant fields.
Here, we provide the weak- and strong-field asymptotics of the Heisenberg-Euler effective Lagrangian at one- and two-loop order. Their explicit derivation is relegated to Appendix~\ref{app:sfa}.
Thereafter, in Sec.~\ref{sec:photonsEFT} we introduce the LCFA and show how it can be employed to construct an effective theory describing the interactions of slowly varying electromagnetic fields and low-frequency photon fields in the QED vacuum.
Here, we mainly concentrate on fluctuation-induced effective interactions at two-loop order.
Finally, we end with conclusions and an outlook in Sec.~\ref{sec:Concls+Outl}.

%%%%%%%%%%%%%%%%%%%%%%%%%%% 
\section{1PI effective action and Heisenberg-Euler effective action -- differences and common ground} \label{sec:HEvs1PI}
%%%%%%%%%%%%%%%%%%%%%%%%%%%

\subsection{Partition function and vacuum persistence amplitude in an external field}

The Heisenberg-Euler action $\Gamma_{\text{HE}}$
\cite{Heisenberg:1935qt} is often viewed as the prototype of an
effective action $\Gamma$, the latter having become a canonized
central object in QFT. Effective actions $\Gamma$ have a precise
meaning as the generating functional of 1PI correlators (proper
vertices) and follow from a standardized QFT construction via the
Legendre transform of the partition function\footnote{In order to keep
  the notation compact, we employ the shorthand forms
  $\int_x\equiv\int{\rm d}^4x$ and $\int_k\equiv\int\frac{{\rm
      d}^4k}{(2\pi)^4}$ for the integrations over position and
  momentum space, respectively. Besides, we simply use $\int$ if the
  integration can be performed in position or momentum space.}
\begin{eqnarray}
Z[J]=\int \mathcal{D} \varphi \, e^{iS[\varphi]+i\int J\varphi}, \label{eq:partfunc}\\
\Rightarrow\quad \Gamma[\phi]=\stackrel[J]{}{\text{sup}}  \Big[ -\int J\phi - i \ln Z[J] \Big]. \label{eq:1PI} 
\end{eqnarray}
Here, $S$ denotes the classical action of the theory to be quantized,
$\varphi$ summarizes the fluctuation fields such as
electrons/positrons and photons in QED, and $J$ is a source that can
be used to generate correlation functions from the partition function.
The above definition~\eqref{eq:1PI} guarantees the 1PI property of $\Gamma$, making it a convenient and elegant
tool for many purposes of QFT.

As it is of particular relevance for understanding our novel results
obtained below, we wish to emphasize that the Heisenberg-Euler action
does not fall into the class of effective actions as constructed from
\Eqref{eq:1PI}. This statement holds both from the perspective of its
historical construction as well as from its modern use in strong-field
physics.

The physical difference is, for instance, apparent from the fact that
the 1PI effective action~\eqref{eq:1PI} depends on the
so-called \textit{classical} field $\phi$. The supremum prescription in
\Eqref{eq:1PI} relates $\phi$ to the expectation value of the
fluctuating quantum field,
\begin{equation}
\phi=\frac{1}{i} \frac{1}{Z[J]} \frac{\delta Z[J]}{\delta J}=\langle\varphi\rangle, \label{eq:defphi}
\end{equation}
(a relation that can be considered both at $J=0$ or for a nonvanishing
source). In turn, the field $\phi$ is obviously the result of a full
quantum averaging process.

By contrast, Heisenberg and Euler \cite{Heisenberg:1935qt} as well as
Weisskopf \cite{Weisskopf} have been interested in the response of the
quantized electron-positron field to a non-quantized \textit{external}
electromagnetic field $\Abar$ which is considered to be given from the
outside. In absence of quantum fluctuations, this external field
would obey an action principle with action $S_{\text{ext}}[\Abar]=
- \frac{1}{4} \int \Fbar_{\mu\nu} \Fbar^{\mu\nu}$, where $\bar
F_{\mu\nu}=\partial_\mu \bar A_\nu-\partial_\nu \bar
A_\mu$. Contrarily, in the presence of quantum fluctuations, the dynamics of the external field  $\Abar$ is governed by the Heisenberg-Euler action,
\begin{equation}
\Gamma_\text{HE}[\Abar]=S_{\text{ext}}[\Abar]+W[\Abar]. \label{eq:GHE}
\end{equation}
The additional contribution $W[\Abar]$ arising from quantum fluctuations has been formalized by Schwinger in terms of the vacuum persistence amplitude
\cite{Schwinger:1951nm},
\begin{equation}
\langle 0_+|0_-\rangle^{\Abar}= e^{iW[\Abar]}, \label{eq:vacper}
\end{equation}
parametrizing the probability amplitude for the vacuum to persist in
the presence of an external field $\Abar$ (``the prescribed
  field'' \cite{Schwinger:1951nm}). The Schwinger functional
$W[\Abar]$ is considered to be a functional of the external field (and
not of a source coupled to a quantum field). It can be written as a
path integral over fluctuating fields,
\begin{equation}
e^{iW[\Abar]}= \int \mathcal{D} q\, e^{i\int \left( -\frac{1}{4}Q_{\mu\nu} Q^{\mu\nu}\right)} e^{iS_\psi[\Abar+q]}, \label{eq:PIW}
\end{equation}
where we employed the shorthand notation
\begin{equation}
e^{i S_{\psi}[\Abar+q]} = \int \mathcal{D}\bar\psi  \mathcal{D} \psi\,
e^{i\int \bar\psi ( -i \slashed{D}[\Abar+q]+m)\psi} . \label{eq:Spsi}
\end{equation}
For a proper comparison with the literature, we point out that our
phase conventions agree with those of \cite{Dittrich:1985yb} and thus
do not include the Maxwell term for the $\Abar$ field in $W[\Abar]$
(contrary to Schwinger's conventions
\cite{Schwinger:1951nm}).\footnote{More precisely, our conventions agree with those of the defining equations (1.45) and (1.48) of \cite{Dittrich:1985yb};
these are slightly different from those of chapter 7 of \cite{Dittrich:1985yb} where $W[\Abar]$ denotes the electron-positron loop.}

In \Eqref{eq:PIW}, we have distinguished between the external
background $\Abar$ and the fluctuating photon field $q$, the latter
being equipped with a kinetic term involving the field strength
$Q_{\mu\nu}=\partial_\mu q_\nu-\partial_\nu q_\mu$. The external field
couples to the fermions $\psi,\bar\psi$, and hence the result of the
path integral depends parametrically on $\Abar$.

We emphasize that \Eqref{eq:PIW} contains no information about the
dynamics that creates $\Abar$ in the first place. This has to be
provided by a separate theory for the external field, which is
conventionally assumed to obey an action principle with action
$S_{\text{ext}}[\Abar]$. In absence of quantum fluctuations,
$\Abar$ would be a solution of this external theory and its
equations of motion given by
\begin{equation} 
\frac{\delta S_{\text{ext}}[\Abar]}{\delta \Abar_\mu}=- \Jbar^{\mu},
\end{equation}
where $\Jbar$ is a classical source for the external field. Upon the
inclusion of quantum fluctuations, the dynamics of $\Abar$ is modified
such that $\Gamma_\text{HE}[\Abar]$ governs the dynamics of the external field.

To one-loop order, the photon fluctuations $\sim \mathcal{D}q$ can be
ignored in \Eqref{eq:PIW} and one obtains the historic answer
\cite{Heisenberg:1935qt,Weisskopf,Schwinger:1951nm}. At
higher loops, starting from two loop on, $W[\Abar]$ also contains
\textit{one-particle reducible} diagrams \cite{Ritus:1975,Ritus:1977,Dittrich:1985yb}, as is obvious from its
definition \eqref{eq:PIW} and will be recalled explicitly
below. Hence, $\Gamma_\text{HE}$ does not correspond to the standard 1PI effective
action.

From a fundamental viewpoint, the concept of a non-quantized external
field $\Abar$ seems somewhat redundant, as the world is fully
quantum. Moreover, a separation into internal and external fields
might seem purely academic. Nevertheless, this concept is perfectly
adjusted to our perception of a real experiment in terms of
classically controlled sources and detectors. In the remainder of this
section, we detail how this useful concept can be extracted from the
full quantum theory.

\subsection{From QED to the Heisenberg-Euler effective action}

In order to develop the formalism, it is useful to envisage a typical
physical system where the external field $\Abar$ is generated by
suitable sources $\Jbar$. The sources (lasers, magnets, etc.) are
macroscopically separated from an interaction region of volume
$V_{\text{I}}$ (focal volume, interaction cavity, etc.). We consider
physical situations where quantum vacuum nonlinearities, i.e.,
higher-order effective couplings of electromagnetic fields mediated by
quantum fluctuations of charged particles, become sizable only within
$V_{\text{I}}$.

Then, the physics inside $V_{\text{I}}$ can create signals (induced
field components, signal photons, etc.) which are ultimately observed
in detectors macroscopically separated from the interaction region
$V_{\text{I}}$. Due to the smallness of the nonlinear effective
couplings among electromagnetic fields induced by quantum fluctuations
of virtual charged particles, the signal may often be of quantum
nature, as it is, e.g., the case for a
single-photon signal to be measured in a
single-photon detector. Still, it is useful to think of the signal as
a contribution to the external field $\Abar$, because it is ultimately
measured far away from the region $V_{\text{I}}$ .

In order to distinguish between applied fields $\Abar_{\text{applied}}$ (e.g., the fields provided by
lasers, or magnets) and the signal
photons $\Abar_{\text{signal}}$, one may decompose the external field as
\begin{equation}
\Abar=\Abar_{\text{applied}}+\Abar_{\text{signal}}. \label{eq:ppsig}
\end{equation}
As the signal $\Abar_{\text{signal}}$ is eventually induced by
$\Abar_{\text{applied}}$, the two components of $\Abar$ will typically
exhibit a causal ordering in time. Similarly, it is possible to
distinguish between the source parts $\Jbar$ that are responsible for
creating $\Abar_{\text{applied}}$ and those that interact with
 $\Abar_{\text{signal}}$ within the detectors.

Now, the quantitative success of classical electrodynamics heuristically implies that the effective self-interactions as well as mutual couplings of $\Abar_{\text{signal}}$ and $\Abar_{\text{applied}}$ mediated by quantum fluctuations $(\psi,\bar\psi,q)$ are dramatically suppressed and essentially vanish outside the interaction volume $V_{\text{I}}$. From the viewpoint of QED, this is a consequence of the locality of the theory and the smallness of its coupling.
This establishes an operational definition of
$\Abar$ in the outside region, where it is related to the sources
$\Jbar$ which control both the creation of $\Abar_{\text{applied}}$ and the
detection of $\Abar_{\text{signal}}$. For the following formalism, it
suffices to just refer to the combined field $\Abar$.
From a conceptual point of view, the details of the choice of
$V_{\text{I}}$ do not really matter. It is the possibility of a
partitioning of the system into an internal interaction and external Maxwellian region that
matters (cf. below).  Correspondingly, there is no need to consider (Casimir-like) effects due to the finite volume of $V_{\text{I}}$:
The interaction volume can always be chosen large enough to render such effects
negligible. In fact, as not even a physical boundary is necessary, the
transition between internal and external regions can be fuzzy.

Apart from the effects of quantum corrections, we expect $\Abar$ to satisfy a classical Maxwell
equation $\delta S_{\text{ext}}/\delta \Abar \simeq -\Jbar$. More
precisely, we assume $\Abar$ to be defined as the solution of the following equation,
\begin{equation}
\partial_\mu \Fbar^{\mu\nu} + C^\nu[\Abar]= -\Jbar^\nu,\label{eq:defC}
\end{equation}
where $C^\nu[\Abar]$ parameterizes quantum corrections which should be negligible in the
outside region, i.e., approximately fulfill $C^\nu[\Abar]=0$ outside $V_\text{I}$. 
By contrast, $C^\nu[\Abar]$ can become relevant in the
interaction region $V_{\text{I}}$, where however $\Jbar=0$. In QED,
$C^\nu[\Abar]$ is perturbatively of $\mathcal{O}(\alpha)$ and
nonlinear and nonlocal in the field, with the nonlinearities and
nonlocalities being controlled by the Compton scale. 

With these prerequisites, let us turn to the standard partition function for QED,
\begin{equation}
Z[J]=\int \mathcal{D} A \, 
e^{i \int \left(-\frac{1}{4} F_{\mu\nu}  F^{\mu\nu} + J_\mu A^\mu\right)}
e^{i S_{\psi}[A]}.
\label{eq:partfuncQED}
\end{equation}
Concentrating on correlation functions of the electromagnetic field, we only include a source term for the gauge field.
Of course, the generalization to sources for the fermions is straightforward.
In a next step, we employ the variable substitution
\begin{equation}
A=\Abar+q,\label{eq:split}
\end{equation}
in order to rewrite \Eqref{eq:partfuncQED} as 
\begin{equation}
Z[J]=e^{i\int \left(-\frac{1}{4} \Fbar_{\mu\nu} \Fbar^{\mu\nu} + J_\mu \Abar^\mu \right)}
\int \mathcal{D} q \, 
e^{i \int  \left[-\frac{1}{4} Q_{\mu\nu}  Q^{\mu\nu}+( \partial_\mu \Fbar^{\mu\nu} + J^\nu)q_\nu\right]}
e^{i S_\psi[\bar A+q]} .
\label{eq:partfuncQED2}
\end{equation}
We emphasize that -- despite its explicit appearance on the right-hand side --
this partition function of course
 does not depend on $\Abar$ but is a functional of the source $J$ only.
Contrary to the standard QFT treatment
where $J$ often plays the role of an auxiliary variable, the source is
needed here to sustain the external field. Still, let us not simply
reduce $J\to\Jbar$, but keep it slightly more general.

As a next step, we \textit{classicalize} the external field $\Abar$:
For this, we assume that the
fluctuation field $q$ only couples to the electron-positron field,
i.e., any direct coupling to the background field should vanish. More precisely, we choose $J$ such that
\begin{equation}
\int d^4x(\partial_\mu \Fbar^{\mu\nu} + J^\nu)q_\nu =0\quad\leftrightarrow\quad J^\nu=-\partial_\mu \Fbar^{\mu\nu}=:-(\partial\bar F)^\nu.
\label{eq:classical}
\end{equation} 
Let us emphasize that for any violation of
\Eqref{eq:classical}, i.e., $\partial_\mu \Fbar^{\mu\nu} +
J^\nu=\mathcal{J}^\nu\neq0$, the remnant source 
$\mathcal{J}^\nu$ could potentially induce a nonvanishing expectation
value $\langle q_\nu \rangle = (1/i Z) (\delta Z/ \delta
\mathcal{J}^\nu)\neq0$. Such an expectation value could mix with $\Abar$ and
thereby lead to inconsistencies 
with the concept of $\Abar$ being an external field.

For sources fulfilling \Eqref{eq:classical}, we have
\begin{equation}
Z[J]\big|_{J=-(\partial\bar F)}=
{\rm e}^{i\int\left(+\frac{1}{4}\Fbar_{\mu\nu}\Fbar^{\mu\nu}\right)}\int \mathcal{D} q \, 
e^{i \int \left( -\frac{1}{4} Q_{\mu\nu}  Q^{\mu\nu}\right) } e^{i S_{\psi}[\Abar+q]} .
\label{eq:partfuncQED3}
\end{equation}
A comparison with Schwinger's vacuum persistence amplitude \eqref{eq:PIW}
shows that 
\begin{equation}
Z[J]\big|_{J=-(\partial\bar F)}={\rm e}^{i\int\left(+\frac{1}{4}\Fbar_{\mu\nu}\Fbar^{\mu\nu}\right)}e^{iW[\Abar]} .
\end{equation}
This suggests introducing the Heisenberg-Euler action by
\begin{align}
\Gamma_{\text{HE}}[\Abar] &:= \biggl(-\int J_\mu\bar  A^\mu -i\ln Z[J] \biggr)\bigg|_{J=-(\partial\bar F)}  \nonumber\\
 &=- \int \frac{1}{4} \Fbar_{\mu\nu} \Fbar^{\mu\nu} + W[\Abar].  \label{eq:defHE}
\end{align}
Note that \Eqref{eq:defHE} does not constitute a Legendre transform, since $J$ is subject to the
constraint \eqref{eq:classical}. 

Since the field $\Abar$ is ultimately created by the classical source $\Jbar$, we demand for 
\begin{equation}
 - \Jbar^\mu \stackrel{\text{!}}{=} \frac{\delta \Gamma_{\text{HE}}[\Abar]}{\delta \Abar_\mu} = \partial_\nu \Fbar^{\nu\mu} + \frac{\delta W[\Abar]}{\delta \Abar_\mu}, \label{eq:HEEoM}
\end{equation}
which implies that the correction term in \Eqref{eq:defC} is given by
$C^\mu[\Abar]= \delta W[\Abar]/\delta \Abar_\mu$.
Hence the correction term can be viewed as a shift in the source term
\begin{equation}
\Jbar\to\Jbar + C[\Abar] = J,
\end{equation}
which is needed in \Eqref{eq:classical} to inhibit that the background
as well as $\Jbar$ provide a source for the fluctuation field $q$.
If we had defined the field $\Fbar$ in terms of the source $\Jbar$ in
combination with the classical field equation $\partial_\mu
\Fbar^{\mu\nu}= -\Jbar^\nu$, we would have arrived at the same
definition \eqref{eq:defHE} for the Heisenberg-Euler action. However,
this definition of the external field would have been inconsistent
with the quantum equation of motion \eqref{eq:HEEoM} from order $\alpha$ on.

This concludes our derivation of the Heisenberg-Euler effective action
$\Gamma_{\text{HE}}$ from the standard QFT partition function of QED.
The result~\eqref{eq:defHE} is in perfect agreement with Schwinger's definition by
means of the vacuum persistence amplitude~\eqref{eq:vacper}.
Our derivation underpins once more that
$\Gamma_{\text{HE}}$ is decisively different from the standard
effective action $\Gamma$, as it also contains one-particle reducible
contributions which contribute to the equations of motion of the
external field.

We end this section with the remark that once $\Gamma_{\text{HE}}$ is
obtained, it can be used for determining $\Abar$ both by purely
classical means or by describing $\Abar$ in terms of a Fock space in a
quantum optical setting. Both treatments of $\Abar$ are useful as well
as legitimate. In particular, it is natural to treat applied
macroscopic fields $\Abar_{\text{applied}}$ classically and the
induced weak signal fields $\Abar_{\text{signal}}$ by means of
Fock space states, as has been suggested in the vacuum emission picture
\cite{Karbstein:2014fva}.

\subsection{Diagrammatic expansion of the Heisenberg-Euler effective action} \label{subsec:diagexp}

\begin{figure}
 \centering
 \includegraphics[width=0.6\columnwidth]{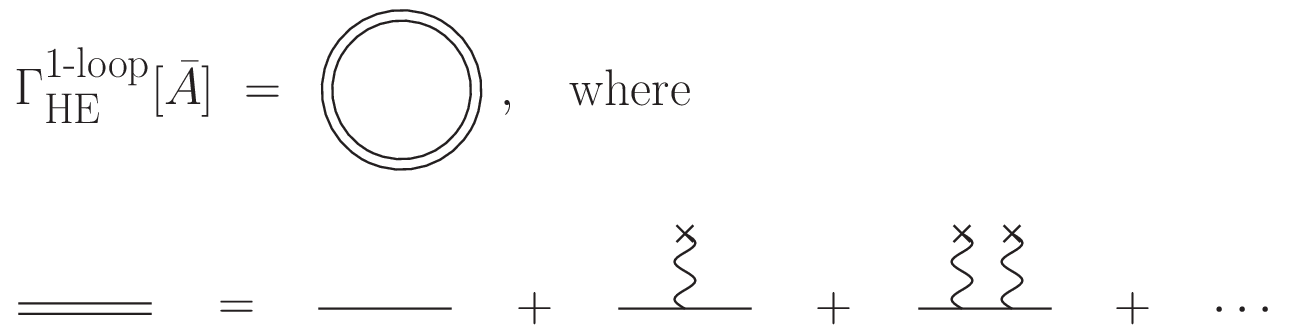}
\caption{Diagrammatic representation of the one-loop Heisenberg-Euler effective action. The double line denotes the dressed fermion propagator accounting for arbitrarily many couplings to the external field $\Abar$, represented by the wiggly lines ending at crosses.}
\label{fig:1loopGamma}
\end{figure}

\begin{figure}
 \centering
 \includegraphics[width=0.55\columnwidth]{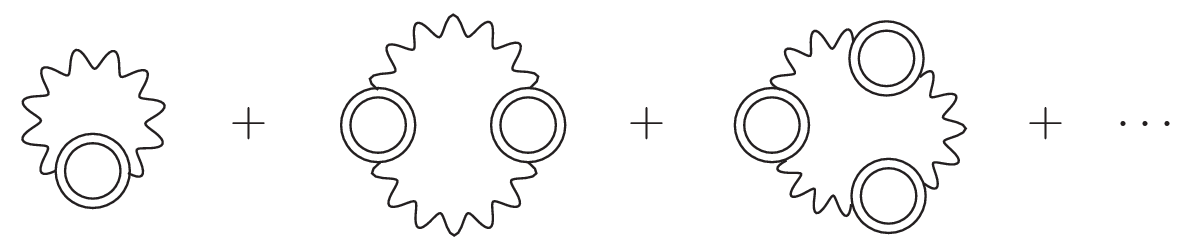}
\caption{Diagrammatic representation of \Eqref{eq:2la}. The wiggly line is the photon propagator; for the definition of the double line, cf. Fig.~\ref{fig:1loopGamma}.}
\label{fig:Rosenkranz}
\end{figure}

Apart from the classical Maxwell term, the Heisenberg-Euler effective
action~\eqref{eq:defHE} is given by the Schwinger functional
$W[\Abar]$, which can be defined in terms of a functional integral,
see \Eqref{eq:PIW}.  The latter encodes quantum corrections giving
rise to effective self-interactions of the external electromagnetic
field; for $\hbar\to 0$ we have $W[\Abar]\to0$. It can be
perturbatively expanded by standard techniques, cf., e.g.,
\cite{Dittrich:1985yb}.  Generically, this expansion can be organized in
  the number of loops,
\begin{equation}
 W[\Abar]=\sum_{l=1}^\infty \Gamma_\text{HE}^{l\text{-loop}}[\Abar] \,,
\end{equation}
with $\Gamma_\text{HE}^{l\text{-loop}}\sim(\frac{\alpha}{\pi})^{l-1}$,
where $\alpha=\frac{e^2}{4\pi}\simeq\frac{1}{137}$ is the
fine-structure constant; we use the Heaviside-Lorentz System with
$c=\hbar=1$.  At each loop order $l$,
$\Gamma_\text{HE}^{l\text{-loop}}=\int_x{\cal
  L}_\text{HE}^{l\text{-loop}}$ accounts for an infinite number of
couplings to the external field, and thus is fully nonperturbative in
the parameter $e\bar A$.
For completeness, we sketch the expansion to two-loop order  in the
following. We begin by noting that the fermionic integral in
\Eqref{eq:Spsi} can be written as a functional determinant,
\begin{equation}
i S_\psi[\Abar+q]= \ln \det \Bigl( -i \slashed{D}[\Abar+q] + m\Bigr). \label{eq:lndet}
\end{equation}
If evaluated at $q=0$, this quantity amounts to the one-loop
Heisenberg-Euler effective action in the external field $\Abar$, i.e.,
$\Gamma_{\text{HE}}^{\text{1-loop}}[\Abar]=S_\psi[\Abar]$; for a
graphical representation, cf. Fig.~\ref{fig:1loopGamma}.
Since $S_\psi$ is a one-loop expression, the two-loop order of
the Schwinger functional is already obtained by performing the
photonic fluctuation integral $\sim \mathcal{D} q$ to Gau\ss{}ian
order.
For this, we expand $S_{\psi}$ about the external field $\Abar$,
\begin{eqnarray}
 S_\psi[\Abar+q]
&=& S_\psi[\Abar] + \int \big(S_\psi^{(1)}[\Abar]\big)^\mu q_\mu + \frac{1}{2} \iint  q_\mu \big(S_\psi^{(2)}[\Abar]\big)^{\mu\nu} q_\nu  + \mathcal{O}(q^3),\label{eq:Spsiexp}
\end{eqnarray}
where we employed the shorthand notation
\begin{equation}
 \big(S_\psi^{(n)}\big)^{\sigma_1\ldots\sigma_n}[\Abar]:= \frac{\delta^n S_\psi[A]}{\delta A_{\sigma_1}\ldots\delta A_{\sigma_n}}\bigg|_{A=\Abar}\,.
\end{equation}
The first-order term corresponds to a one-loop photon current induced
by the field $\Abar$, and the Hessian is related to the one-loop
photon polarization tensor
$\Pi^{\mu\nu}[\Abar]:=\big(S_\psi^{(2)}\big)^{\mu\nu}[\Abar]$
  evaluated in the external field $\Abar$; for completeness note that
  this definition of the photon polarization tensor differs from that
  of \cite{Karbstein:2015cpa} by an overall minus sign.
To Gau\ss{}ian order, we ignore
the terms of $\mathcal{O}(q^3)$ in the exponent, resulting in
\begin{equation}
e^{iW[\Abar]}\simeq e^{i S_\psi[\Abar]} \int \mathcal{D}q\, e^{i \int \big(S_\psi^{(1)}[\Abar]\big)^\mu q_\mu 
- \frac{i}{2}  \iint  q_\mu \big(D^{-1} - \Pi[\Abar]\big)^{\mu\nu} q_\nu 
}.
\label{eq:W1}
\end{equation}
In principle, terms of ${\cal O}(q^3)$ in the exponent can, of course, be treated perturbatively to any desired order.
The quantity $ \bigl(D^{-1}\bigr)^{\mu\nu}$ arises from the Maxwell
term for the fluctuations and denotes the inverse photon
propagator. E.g., in momentum space and accounting for a gauge-fixing term
(generalized Lorenz gauge), we have
\begin{equation}
  D^{\mu\nu}(p) = \frac{1}{p^2 -i \epsilon} \left( g^{\mu\nu} - (1-\xi) \frac{p^\mu p^\nu}{p^2-i\epsilon} \right),
\label{eq:photprop}
\end{equation}
where $\xi=1$ in the Feynman gauge.
Performing the integration over $q$ in \Eqref{eq:W1}, we arrive at
\begin{eqnarray}
e^{iW[\Abar]}&\simeq& e^{i S_\psi[\Abar]} 
e^{\frac{i}{2} \iint  \big(S_\psi^{(1)}[\Abar]\big)_\mu  \Big[\big(D^{-1} - \Pi[\Abar]\big)^{-1}\Big]^{\mu\nu}  \big(S_\psi^{(1)}[\Abar]\big)_\nu}
\,\det{}^{-1/2} \big(D^{-1}-\Pi[\Abar]\big). \label{eq:W2}
\end{eqnarray}
To Gau\ss{}ian order in the photon fluctuations, we thus obtain for the Schwinger functional
\begin{equation}
  W[\Abar] \simeq S_\psi[\Abar] - \frac{1}{2} \ln \det  \big(D^{-1}-\Pi[\Abar]\big) + \frac{1}{2}  \iint  \big(S_\psi^{(1)}[\Abar]\big)_\mu  \Big[\big(D^{-1} - \Pi[\Abar]\big)^{-1}\Big]^{\mu\nu}   \big(S_\psi^{(1)}[\Abar]\big)_\nu.
\label{eq:W3}
\end{equation}
The first term on the right-hand side corresponds to the one-loop contribution to the
Heisenberg-Euler effective action (called $W^{(1)}[\Abar]$ in
\cite{Schwinger:1951nm,Dittrich:1985yb}). The other two terms contain
the complete two-loop order contribution as well as subclasses of diagrams to
arbitrarily high loop order. To make this manifest, we expand the $\ln \det$ term as follows,
\begin{equation}
  \ln\det \!\big(D^{-1}-\Pi[\Abar]\big) = \Tr\ln (1-D\Pi[\Abar]) + \Tr\ln D^{-1} = -\Tr (D\Pi) + \frac{1}{2}\Tr(D\Pi D\Pi) + \mathcal{O}(\Pi^3),\!
\label{eq:2la}
\end{equation}
where in the last step, we have dropped field-independent 
constants. The $\Tr(D\Pi)$ term corresponds exactly to the two-loop
contribution to the Heisenberg-Euler action that has first been
computed in \cite{Ritus:1975}; see also
\cite{Dittrich:1985yb,Fliegner:1997ra,Kors:1998ew,Dunne:2004nc}. 
This contribution as well as all higher order terms in \Eqref{eq:2la} are one-particle irreducible from a diagrammatic viewpoint; see Fig.~\ref{fig:Rosenkranz}.

\begin{figure}
 \centering
 \includegraphics[width=0.72\columnwidth]{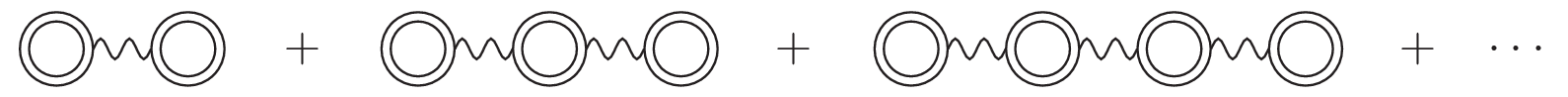}
\caption{One-particle reducible diagrams constituting the last term in \Eqref{eq:W2}. For the definition of the double line, cf. Fig.~\ref{fig:1loopGamma}.}
\label{fig:RedKette}
\end{figure}

The last term in \Eqref{eq:W2}, however, contains 
\begin{equation}
\big(D^{-1} - \Pi[\Abar]\big)^{-1}=D-D\Pi D + D\Pi D \Pi D+ \dots,\label{eq:propexp}
\end{equation}
corresponding to the Dyson series of the full one-loop resummed photon
propagator. In the last term of \Eqref{eq:W2}, this resummed
propagator interconnects two one-loop photon currents $\sim
\big(S_\psi^{(1)}[\Abar]\big)$. All the diagrams arising when adopting
the expansion~\eqref{eq:propexp} in the last term in \Eqref{eq:W2} are
one-particle reducible; see Fig.~\ref{fig:RedKette}.

In turn, the two-loop Heisenberg-Euler effective action consists of a 1PI and a 1PR diagram and is given by
\begin{equation}
 \Gamma_{\text{HE}}^{\text{2-loop}}[\Abar]=\underbrace{\frac{1}{2}\Tr(D\Pi[\Abar])}_{=:\Gamma_{\text{HE}}^{\text{2-loop}}\big|_{1{\rm PI}}} +\underbrace{\frac{1}{2}  \iint  \big(S_\psi^{(1)}[\Abar]\big)_\mu  D^{\mu\nu}   \big(S_\psi^{(1)}[\Abar]\big)_\nu}_{=:\Gamma_{\text{HE}}^{\text{2-loop}}\big|_{1{\rm PR}}} \,. \label{eqGEH2l}
\end{equation}
\begin{figure}
 \centering
 \includegraphics[width=0.43\columnwidth]{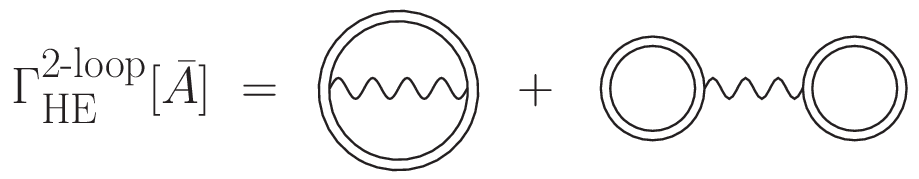}
\caption{Diagrams constituting the two-loop Heisenberg-Euler effective action. Obviously, we have $\Gamma_{\text{HE}}^{\text{2-loop}}=\Gamma_{\text{HE}}^{\text{2-loop}}\big|_{1{\rm PI}}+\Gamma_{\text{HE}}^{\text{2-loop}}\big|_{1{\rm PR}}$.
Note that the first diagram amounts to the leftmost one in Fig.~\ref{fig:Rosenkranz}, where it is drawn in a slightly different way; for the definition of the double line, cf. Fig.~\ref{fig:1loopGamma}.}
\label{fig:2loopGamma}
\end{figure}

The existence as a matter of principle of the 1PR term in
\Eqref{eqGEH2l} has been known for a long time. It has, however,
been argued that this term vanishes for constant external fields
\cite{Ritus:1975,Dittrich:1985yb}. Let us reproduce this argument
for reasons of completeness: a crucial building block of the 1PR
term is $\big(S_\psi^{(1)}[\Abar]\big)^\mu$, which corresponds to the
one-loop photon current which will be called $j_{1\text{-loop}}^\mu[\Abar]$
below. For a constant external field, $j_{1\text{-loop}}^\mu[\Abar]$ does not
depend on any spacetime point $x$ either. On the other hand, $j_{1\text{-loop}}^\mu[\Abar]$
is a Lorentz 4-vector. The vector index of the current can only be
generated from the building blocks $\bar F$, $\partial$ and $x$.
However, for constant fields $\partial^\mu F^{\nu\kappa}=0$
and for an $x$ independent current, all conceivable combinations
with one vector index vanish and so does the current (an explicit
verification of this fact in momentum space is given below). 

While this part of the argument holds true in the full analysis,
it does not necessarily imply that the 1PR diagram in Fig.~\ref{fig:2loopGamma} vanishes.
In fact, the two currents in the 1PR diagram are convoluted with a
photon propagator, describing a long-range force with an IR
singularity $\sim 1/p^2$ in the propagator. Hence, it is a 
quantitative question as to whether the currents approaching zero
are outbalanced by the IR singularity of the photon propagator.
In the subsequent sections, we give proof that the result is finite.

Heuristically, it is clear that the above-mentioned argument for
the vanishing of the current no longer holds as soon as the external
field supports a slightest inhomogeneity somewhere in spacetime.
So, $j^\mu[\Abar]=0$ strictly relies on $\partial^\mu F^{\nu\kappa}=0$ for all $x$.
On the other hand, the existence of massless long-range
fluctuations in QED is independent of the constant-field
assumption. Therefore, the 1PR term is expected to be finite for any
realistic  field.

%%%%%%%%%%%%%%%%%%%%%%%%%%% 
\section{The Heisenberg-Euler effective action in constant electromagnetic fields} \label{sec:knownHE}
%%%%%%%%%%%%%%%%%%%%%%%%%%%

In the following, we first summarize some of our explicit
results for the Heisenberg-Euler effective action, concentrating on
fully analytic expressions in asymptotic field-strength limits for spinor QED. This
provides for a first glance at the parametric dependence of the
various contributions, and elucidates the regime of relevance of the
two-loop 1PR term. Details of the calculations are deferred to the
subsequent sections.

For constant external electromagnetic fields, 
$\bar F^{\mu\nu}=\text{const.}$, Lorentz and gauge invariance constrain
$\Gamma_\text{HE}$ to depend on $\Abar$ only in terms of the two
scalar invariants \cite{Euler:1935zz,Heisenberg:1935qt} ${\cal
  F}=\frac{1}{4}\bar F_{\mu\nu}\bar
F^{\mu\nu}=\frac{1}{2}\bigl(\vec{B}^2-\vec{E}^2\bigr)$ and ${\cal
  G}=\frac{1}{4}\bar F_{\mu\nu}{}^*\bar
F^{\mu\nu}=-\vec{B}\cdot\vec{E}$, with dual field strength tensor
$^*\bar F^{\mu\nu}=\frac{1}{2}\epsilon^{\mu\nu\alpha\beta}\bar
F_{\alpha\beta}$. Here, $\epsilon^{\mu\nu\alpha\beta}$ is the totally
antisymmetric tensor ($\epsilon^{0123}=1$), and our metric convention
is $g_{\mu \nu}=\mathrm{diag}(-1,+1,+1,+1)$. 
In addition, CP invariance of QED dictates $\Gamma_\text{HE}$ to be even in the
pseudoscalar quantity ${\cal G}$, $\Gamma_\text{HE}[\Abar]= \Gamma_\text{HE}({\cal F},{\cal G}^2)$. 
To keep notations compact, we also introduce the
dimensionless quantities $\tilde\F=(\frac{e}{m^2})^2\F$ and
$\tilde\G=(\frac{e}{m^2})^2\G$.  Finally note that the action and the
Lagrangian are trivially related in constant fields, differing only by an overall 
volume factor, i.e., $\Gamma_\text{HE}={\cal
  L}_\text{HE}\int_x$.

In constant external fields, ${\cal L}_\text{HE}^{1\text{-loop}}$ and ${\cal L}_\text{HE}^{2\text{-loop}}\big|_{1{\rm PI}}$ are known explicitly in terms of parameter integral representations for both spinor \cite{Heisenberg:1935qt,Ritus:1975} and scalar \cite{Weisskopf,Ritus:1977} QED; cf. \cite{Dunne:2004nc} for a review.
For instance, the on-shell renormalized one-loop effective Lagrangian for spinor QED is given by \cite{Heisenberg:1935qt,Schwinger:1951nm},
\begin{equation}
 {\cal L}_\text{HE}^{1\text{-loop}}=-\frac{1}{8\pi^2}\int_{0}^{\infty}\frac{{\rm d}T}{T^3}\,{\rm e}^{-m^2T}\biggl\{\frac{(e\epsilon T)(e\eta T)}{\tan(e\epsilon T)\tanh(e\eta T)}-\frac{2}{3}(eT)^2{\cal F}-1\biggr\},
\end{equation}
where $\epsilon=\bigl(\sqrt{{\cal F}^2+{\cal G}^2}-{\cal F}\bigr)^{1/2}$ and $\eta=\bigl(\sqrt{{\cal F}^2+{\cal G}^2}+{\cal F}\bigr)^{1/2} $ are the secular invariants in constant electromagnetic fields. The analogous expression for
 ${\cal L}_\text{HE}^{2\text{-loop}}\big|_{1{\rm PI}}$ is given in \Eqref{eq:L2} in the appendix for spinor QED. For completeness, we also note that mass renormalization has to be taken into account from two loops on for diagrams involving fermion loops with internal radiative corrections; see, e.g., \cite{Fliegner:1997ra}.

As indicated above and determined explicitly below, the 1PR contribution ${\cal L}_\text{HE}^{2\text{-loop}}\big|_{1{\rm PR}}$ depicted in Fig.~\ref{fig:2loopGamma} is finite also in the constant field limit.
Based on the structure of the LCFA, we detail below how
the exact expression for ${\cal L}_\text{HE}^{2\text{-loop}}\big|_{1{\rm PR}}$ in constant fields can 
be inferred from the constant-field result for ${\cal L}_\text{HE}^{1\text{-loop}}$, yielding
\begin{align}
{\cal L}_\text{HE}^{2\text{-loop}}\big|_{1{\rm PR}}
&= \frac{1}2{}\frac{\partial{\cal L}_\text{HE}^{1\text{-loop}}}{\partial F^{\mu\nu}}\frac{\partial{\cal L}_\text{HE}^{1\text{-loop}}}{\partial F_{\mu\nu}} \nonumber\\
&= \frac{1}{2}{\cal F}\biggl[\Bigl(\frac{\partial{\cal L}_\text{HE}^{1\text{-loop}}}{\partial{\cal F}}\Bigr)^2
-  \Bigl(\frac{\partial{\cal L}_\text{HE}^{1\text{-loop}}}{\partial{\cal G}}\Bigr)^2\biggr]
 + {\cal G}\,\frac{\partial{\cal L}_\text{HE}^{1\text{-loop}}}{\partial{\cal F}}\,\frac{\partial{\cal L}_\text{HE}^{1\text{-loop}}}{\partial{\cal G}} \,. \label{eq:1PR}
\end{align}
In turn, ${\cal L}_\text{HE}^{2\text{-loop}}\big|_{1{\rm PR}}$ can be expressed in terms of a double parameter integral. 

For illustration, let us concentrate on the weak and strong field asymptotics of
${\cal L}_\text{HE}^{1\text{-loop}}$ and ${\cal L}_\text{HE}^{2\text{-loop}}={\cal L}_\text{HE}^{2\text{-loop}}\big|_\text{1PI}+{\cal L}_\text{HE}^{2\text{-loop}}\big|_\text{1PR}$
for spinor QED.
In the weak field limit, characterized by $\{\tilde\F,\tilde\G\}\ll1$, the well-known literature results read \cite{Euler:1935zz,Heisenberg:1935qt,Weisskopf},
\begin{equation}
 \frac{{\cal L}_\text{HE}^{1\text{-loop}}}{m^4}=\frac{1}{4\pi^2}\frac{1}{90}\Bigl[(4\tilde{\cal F}^2+7\tilde{\cal G}^2)-\tilde\F\Bigl(\frac{32}{7}\tilde\F^2+\frac{52}{7}\tilde\G^2\Bigr)+{\cal O}(\epsilon^8)\Bigr],
 \label{eq:L1loopweakfield}
\end{equation}
and \cite{Ritus:1975}
\begin{equation}
 \frac{{\cal L}_\text{HE}^{2\text{-loop}}\big|_{1{\rm PI}}}{m^4} = \frac{\alpha}{\pi}\frac{1}{4\pi^2}\frac{1}{90}
 \Bigl[\Bigl(\frac{160}{9}\tilde{\cal F}^2+\frac{1315}{36}\tilde{\cal G}^2\Bigr)
 -\tilde\F\Bigl(\frac{1219}{45}\tilde{\cal F}^2+\frac{2164}{45}\tilde{\cal G}^2\Bigr)+{\cal O}(\epsilon^8)\Bigr],
 \label{eq:L2loopweakfield}
\end{equation}
where we count ${\cal O}(\frac{e\bar F^{\mu\nu}}{m^2})\sim{\cal O}(\epsilon)$. 
The terms given explicitly in Eqs.~\eqref{eq:L1loopweakfield} and \eqref{eq:L2loopweakfield} amount to the 1PI diagrams depicted in Figs.~\ref{fig:1loopGamma} and \ref{fig:2loopGamma} with the fermion loop featuring four and six couplings to the external field, respectively.
For the two-loop 1PR contribution, we obtain from \Eqref{eq:1PR} the new result
\begin{equation}
\frac{{\cal L}_\text{HE}^{2\text{-loop}}\big|_{1{\rm PR}}}{m^4}
 = \frac{\alpha}{\pi} \frac{1}{4\pi^2}\frac{1}{180} \Bigl[\tilde{\cal F}\Bigl(\frac{32}{45}\tilde{\cal F}^2+\frac{14}{45}\tilde{\cal G}^2\Bigr) + {\cal O}\bigl(\epsilon^8\bigr)\Bigr] . \label{eq:1PRwf}
\end{equation}
The contribution given explicitly here stems from the 1PR diagram in Fig.~\ref{fig:2loopGamma} with each fermion loop
exhibiting three couplings to the external field.
For $|\tilde\F|\gg1$ and $|\tilde\G|\ll1$ corresponding to
the cases of strong electric or magnetic fields, we obtain (for the derivation, see App.~\ref{app:sfa})
\begin{equation}
 \frac{{\cal L}_\text{HE}^{1\text{-loop}}}{m^4}=\frac{1}{4\pi^2}\frac{1}{3}\biggl\{{\tilde\F}
 \,\Bigl[\ln\sqrt{\tilde\F}
 +{\cal O}\bigl((\tfrac{1}{\tilde\F})^0\bigr)\Bigr]
 +\frac{1}{2\sqrt{2}}\frac{\tilde\G^2}{\tilde\F}
 \Bigl[\sqrt{{\tilde\F}} +{\cal O}\bigl((\tfrac{1}{\tilde\F})^0\bigr)\Bigr] + {\cal O}(\tilde\G^4)
 \biggr\}\,,
 \label{eq:L1loopstrongfield}
\end{equation}
and
\begin{equation}
 \frac{{\cal L}_\text{HE}^{2\text{-loop}}\big|_{1{\rm PI}}}{m^4}=\frac{\alpha}{\pi}\frac{1}{4\pi^2}\frac{1}{4}\biggl\{{\tilde\F}
 \,\Bigl[\ln\sqrt{\tilde\F}
 +{\cal O}\bigl((\tfrac{1}{\tilde\F})^0\bigr)\Bigr]
 -\frac{1}{3\sqrt{2}}\frac{\tilde\G^2}{\tilde\F}
 \Bigl[\sqrt{{\tilde\F}} +{\cal O}\bigl((\tfrac{1}{\tilde\F})^0\bigr)\Bigr] + {\cal O}(\tilde\G^4)
 \biggr\}\,,
 \label{eq:L2loopstrongfield}
\end{equation}
where
\begin{equation}
 \sqrt{\tilde\F} = \sqrt{|\tilde\F|}\, \Bigl\{\Theta(\tilde\F)- i \Theta(-\tilde\F)\Bigr\} \,. \label{eq:tF}
\end{equation}
In addition to the well-known \textit{leading-log} terms \cite{Ritus:1975,Ritus:1977,Dittrich:1985yb,Dunne:2002ta}, Eqs.~\eqref{eq:L1loopstrongfield} and \eqref{eq:L2loopstrongfield}
also account for the strongly suppressed contribution $\sim\tilde\G^2$ which is of relevance for the photon polarization tensor (cf. Sec.~\ref{sec:softphotonprop} below).
Note that this contribution is suppressed as $\sim\tilde\G^2/\sqrt{\tilde\F}$, such that the criterion $|\tilde\G|\ll1$ imposed for the expansion
seems actually rather conservative, and it might be sufficient to demand $\tilde\G^2/\sqrt{\tilde\F}\ll1$ instead. However, we have not analyzed the scaling of any terms at ${\cal O}(\tilde\G^4)$.
Apart from an overall parametric suppression of ${\cal L}_\text{HE}^{2\text{-loop}}\big|_{1{\rm PI}}$ by a factor of $\frac{\alpha}{\pi}$, the weak and strong field limits of ${\cal L}_{\rm HE}^{1\text{-loop}}$ and ${\cal L}_\text{HE}^{2\text{-loop}}\big|_{1{\rm PI}}$ are of the same structure and only differ in the specific numerical coefficients.
By contrast, the 1PR contribution to ${\cal L}_\text{HE}^{2\text{-loop}}$ scales as 
\begin{equation}
 \frac{{\cal L}_\text{HE}^{2\text{-loop}}\big|_{1{\rm PR}}}{m^4}=\frac{\alpha}{\pi}\frac{1}{4\pi^2}\frac{1}{6}\biggl\{{\tilde\F}
 \,\biggl[\frac{1}{3}\ln^2\sqrt{\tilde\F}-\Bigl(\frac{1}{3}-8\zeta'(-1)\Bigr)\ln\sqrt{\tilde\F} +{\cal O}\bigl((\tfrac{1}{\tilde\F})^0\bigr)\biggr] + {\cal O}(\tilde\G^2)
 \biggr\}\,, \label{eq:1PRsf}
\end{equation}
from which we infer $({\cal L}_\text{HE}^{2\text{-loop}}\big|_{1{\rm PR}})/({\cal L}_\text{HE}^{2\text{-loop}}\big|_{1{\rm PI}})\sim\frac{2}{9}\ln\sqrt{\tilde\F}$,
implying that ${\cal L}_\text{HE}^{2\text{-loop}}\big|_{1{\rm PR}}$ dominates over ${\cal L}_\text{HE}^{2\text{-loop}}\big|_{1{\rm PI}}$ in this limit.
This dominance due to the occurrence of a squared logarithm is a direct consequence of the 1PR structure.
For completeness, also note that ${\cal L}_\text{HE}^{2\text{-loop}}/{\cal L}_\text{HE}^{1\text{-loop}}\sim\frac{1}{6}\frac{\alpha}{\pi}\ln\sqrt{\tilde\F}$. The criterion of apparent convergence of the loop expansion hence suggests the breakdown of the perturbative loop expansion for the Heisenberg-Euler action at exponentially large fields.

Apart from these constant-field results, only a few exact results for
$\Gamma_\text{HE}^{1\text{-loop}}$ in specific (one-dimensional) field
inhomogeneities are known explicitly; cf., e.g.,
\cite{Cangemi:1995ee,Dunne:1997kw,Dunne:1998ni,Kim:2009pg},
and
\cite{Dunne:2004nc} for a review.  Also note that the effective action
vanishes identically for the case of a single monochromatic plane wave field \cite{Schwinger:1951nm}.
On the three-loop level, first analytical results for the 1PI part of $\Gamma_\text{HE}^{3\text{-loop}}$ have been obtained in 1+1 dimensions \cite{Huet:2009cy,Huet:2011kd}.
No further analytical results for $\Gamma_\text{HE}^{l\text{-loop}}$ with $l>2$
as well as for more-dimensional field inhomogeneities are available so far.

%%%%%%%%%%%%%%%%%%%%%%%%%%% 
\section{Effective theory of low-frequency photons in slowly varying electromagnetic fields} \label{sec:photonsEFT}
%%%%%%%%%%%%%%%%%%%%%%%%%%%

\subsection{Locally constant field approximation}

In the spirit of the LCFA, the Heisenberg-Euler effective action for constant fields can also be adopted for slowly varying inhomogeneous fields.
The LCFA amounts to substituting $\bar F^{\mu\nu}\to \bar F^{\mu\nu}(x)$  in the constant-field result for the Lagrangian, such that ${\cal L}_\text{HE}({\cal F},{\cal G}^2)\to{\cal L}_\text{HE}\bigl({\cal F}(x),{\cal G}^2(x)\bigr)$. 
In turn, the corresponding action becomes a functional of a varying field $\bar F^{\mu\nu}(x)$, i.e., $\Gamma_\text{HE}\bigl[{\cal F}(x),{\cal G}^2(x)\bigr]=\int_x {\cal L}_\text{HE}\bigl({\cal F}(x),{\cal G}^2(x)\bigr)$.

The deviations of this LCFA result from the corresponding -- typically unknown -- exact result for $\Gamma_\text{HE}$ in the particular inhomogeneous background field profile under consideration are
of order ${\cal O}\bigl((\tfrac{\upsilon}{m})^2\bigr)$, where $\upsilon$ delimits the moduli of the frequency and momentum components of the considered inhomogeneous field from above  \cite{Galtsov:1982,Karbstein:2015cpa}.
The reasoning to arrive at this conclusion is as follows: As $\Gamma_\text{HE}$ is both a Lorentz scalar and a gauge invariant quantity, and the associated Lagrangian should be ``almost local'' for slowly varying fields,
its dependence on the external field $\Abar^\mu(x)$ should be expressible in terms of $\bar F^{\mu\nu}(x)$, ${}^*\bar F^{\mu\nu}(x)$ and derivatives thereof.
Any scalar quantity made up of combinations of $\bar F$, ${}^*\bar F$ and $\partial$ is necessarily even in $\partial$. Canonical power-counting implies that the occurrence of any derivative $\partial$ has to be balanced by a dimensionful scale. In QED and for generic laboratory fields, this scale is provided by the electron mass $m$, leading to the above criterion.
This implies that the LCFA constitutes a good approximation for inhomogeneous fields fulfilling $\upsilon\ll m$. In position space this criterion translates to the requirement that the inhomogeneous fields under consideration should only vary on scales much larger than the Compton wavelength $\lambdabar_{\rm C}$ and time $\lambdabar_{\rm C}/c$ of the electron; cf. Sec.~\ref{sec:intro}. Explicit results for higher orders in the derivative expansion show, that the dimensional balancing of derivatives can also be taken over by the field strength itself for strong fields $|e\bar F(x)|\gg m^2$, thereby increasing the validity range of the LCFA in that regime \cite{Gusynin:1998bt}.

\subsection{Effective action for low-frequency photons}

In a next step, we employ the LCFA result for the 1PI part of $\Gamma_\text{HE}$ as an effective action $\Gamma_{\rm eff}$, describing the
propagation and interactions of dynamical low-frequency photon fields
in the quantum vacuum subject to the slowly varying
external field.  More precisely, we define this
effective action as
\begin{equation}
 \Gamma_{\rm eff}\bigl[a(x),\bar F(x)\bigr]:=-\frac{1}{4}\int_x f_{\mu\nu}f^{\mu\nu}+\underbrace{\biggl(\Gamma_\text{HE}\bigl[{\cal F}(x),{\cal G}^2(x)\bigr]\big|_{1{\rm PI}}+\int_x{\cal F}(x)\biggr)\bigg|_{\bar F\to \bar F+f}}_{=: \Gamma_{\rm int}[a(x),\bar F(x)]} \,, \label{eq:Gammaeff}
\end{equation}
where $\Gamma_\text{HE}\big|_{1{\rm PI}}$ denotes the 1PI part of the Heisenberg-Euler effective action,
and the field strength tensor $\bar F$ is understood to be shifted as follows \cite{BialynickaBirula:1970vy,Galtsov:1982,Karbstein:2015cpa},
\begin{equation}
 \bar F(x)\to \bar F(x)+f(x)\,.
\end{equation}
After this shift, $\bar F(x)$ describes the slowly varying external
field with $\upsilon\ll m$, and $f^{\mu\nu}(x)=\partial^\mu a^\nu(x) -
\partial^\nu a^\mu(x)$ is to be interpreted as the field strength
tensor of a dynamical photon field $a^\mu(x)=\int_p {e}^{ipx}
a^\mu(p)$, with $a^\mu(p)$ receiving all its relevant contributions
from the momentum regime where $\{|p^0|,|\vec{p}|\}\lesssim\upsilon\ll
m$. In many cases of physical interest, $\bar F(x)$ plays the role of
the applied field and $a(x)$ that of a signal field as introduced in
\Eqref{eq:ppsig}.

It is then convenient to organize $\Gamma_{\rm int}$ in terms of interactions involving $n\in\mathbb{N}_0$ photon fields, i.e., $\Gamma_{\rm int}=\sum_{n=0}^\infty \Gamma^{(n)}_{\rm int}$, with $\Gamma^{(n)}_{\rm int}\equiv \Gamma^{(n)}_{\rm int}\bigl[a(x),{\cal F}(x),{\cal G}^2(x)\bigr]\sim a^n$.
\begin{figure}
 \centering
 \includegraphics[width=0.74\columnwidth]{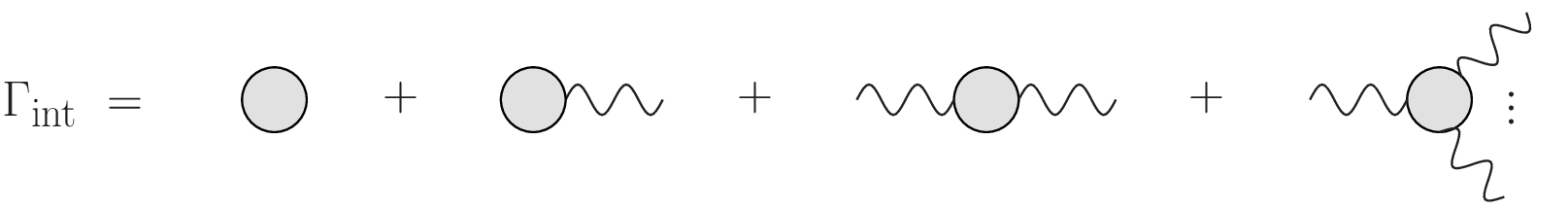}
\caption{Graphical representation of $\Gamma_{\rm int}=\sum_{n=0}^\infty \Gamma^{(n)}_{\rm int}$. The gray bubbles symbolize the effective couplings of $n$ low-frequency photon fields $a(x)$; generically they are made up of 1PI diagrams of arbitrary loop order. In momentum space they are given by $s_{(n)}^{\sigma_1\ldots\sigma_n}(k_1,\ldots, k_n)$ defined in \Eqref{eq:s}.}
\label{fig:S_int}
\end{figure}
For a graphical representation of this expansion, cf. Fig.~\ref{fig:S_int}.
More specifically, we have
\begin{equation}
 \Gamma^{(n)}_{\rm int}=\frac{1}{n!}\int_x\,\prod_{j=1}^n\biggl(f^{\mu_j\nu_j}(x)\frac{\partial}{\partial\bar F^{\mu_j\nu_j}(x)}\biggr)\Bigl({\cal L}_{\rm HE}\bigl({\cal F}(x),{\cal G}^2(x)\bigr)\big|_{1{\rm PI}}+{\cal F}(x)\Bigr) ,
\end{equation}
which implies that ${\cal L}_{\rm HE}\bigl({\cal F}(x),{\cal G}^2(x)\bigr)\big|_{1{\rm PI}}$ generates effective photon interactions to any order in $n$ \cite{Galtsov:1982,Martin:2003gb}.
As the LCFA results in a local Lagrangian, by construction all these effective interactions are local with respect to $f^{\mu\nu}(x)$, and correspondingly
in $a^\mu(x)$.
Let us also emphasize that by construction we have to assume that the combination of any given number $n$ of low-frequency photons again gives rise to a low-frequency photon.
For the following discussion it is more convenient to turn to momentum space where the derivatives acting on the photon fields contained in $f^{\mu\nu}(x)$ translate into multiplicative factors of the associated momenta, i.e.,
$f^{\mu\nu}(x)= i\int_p {e}^{ipx} a_\sigma(p)\bigl[p^\mu g^{\nu\sigma} - p^\nu g^{\mu\sigma}\bigr]$.
This results in
\begin{equation}
 \Gamma^{(n)}_{\rm int}=\frac{1}{n!}\prod_{j=1}^n\biggl(\int_{p_j} a_{\sigma_j}(p_j)\biggr) s_{(n)}^{\sigma_1\ldots\sigma_n}(p_1,\ldots, p_n) , \label{eq:S_n}
\end{equation}
with the effective $n$-photon couplings $s_{(n)}^{\sigma_1\ldots\sigma_n}(p_1,\ldots,p_n)$ (1PI proper vertices) given by
\begin{multline}
 s_{(n)}^{\sigma_1\ldots\sigma_n}(p_1,\ldots, p_n) \\
 =(2i)^n \int_x{e}^{ix\sum_{j=1}^n p_j}\prod_{j=1}^n\biggl(p_j^{\mu_j} g^{\nu_j\sigma_j}
 \frac{\partial}{\partial\bar F^{\mu_j\nu_j}(x)}\biggr) \Bigl({\cal L}_{\rm HE}\bigl({\cal F}(x),{\cal G}^2(x)\bigr)\big|_{1{\rm PI}}+{\cal F}(x)\Bigr) . \label{eq:s}
\end{multline}
The latter obviously fulfill the Ward identity $(p_j)_{\sigma_j}s_{(n)}^{\sigma_1\ldots\sigma_j\ldots\sigma_n}(p_1,\ldots,p_j,\ldots,p_n)=0$ for any fixed value of $1\leq j \leq n$.

For $\bar F={\rm const}.$ the external field cannot absorb or supply momentum, and the $x$ integration in \Eqref{eq:s} can be performed right away, resulting in an overall delta function, $\int_x{e}^{ix\sum_{j=1}^n p_j}=(2\pi)^4\delta\bigl(\sum_{j=1}^n p_j^\mu\bigr)$,
ensuring four-momentum conservation in the effective coupling of $n$ photons.
Hence, in this limit the effective $n$ photon interactions are of the same momentum structure as at zero external field.
However, for $\bar F={\rm const}.\neq0$, also effective couplings involving an odd number of photons are induced.
This is in contrast to the zero-field case, where
fermion loops with an odd number of photon couplings of course vanish identically because of Furry's theorem.

The contribution $s_{(1)}^{\sigma_1}(p_1)$ in~\Eqref{eq:s} constitutes a photon current \cite{Galtsov:1971xm,Karbstein:2014fva} and $s_{(2)}^{\sigma_1\sigma_2}(p_1,p_2)$ a photon polarization tensor \cite{Karbstein:2015cpa}.
In more conventional notations, the quantum corrections to the effective action up to quadratic order in $a^\mu$ are given by
\begin{equation}
\Gamma_{\rm int}= \Gamma_{\rm int}^{(0)}+\int_p a_\sigma(p)j^\sigma(p)+\frac{1}{2}\int_p\int_{p'}a_\rho(p)s_{(2)}^{\rho\sigma}(p,p')a_\sigma(p') +{\cal O}(a^3), \label{eq:S}
\end{equation}
with $j^\sigma(p):= s_{(1)}^{\sigma}(p)$.
The neglected higher-order terms of ${\cal O}(a^3)$ correspond to effective interactions involving three or more photons, giving rise to, e.g., direct light-by-light scattering \cite{Euler:1935zz,Karplus:1950zz},
photon splitting \cite{BialynickaBirula:1970vy,Adler:1971wn,Papanyan:1971cv} and higher-harmonic generation \cite{Bhartia:1978,Bhartia:1980,DiPiazza:2005jc,Fedotov:2006ii}.

Obviously no real (on-shell) photons can be generated from constant external fields, as
\begin{equation}
 j^\sigma(p)\big|_{\bar F={\rm const}.}\sim\int_x {e}^{ixp}p^\sigma=(2\pi)^4\delta(p)p^\sigma . \label{eq:jgleich0?}
\end{equation}
The physical reason for this is that a constant external field cannot supply momentum to the virtual charged particle-antiparticle fluctuations.
Still, the fields $a^\mu$ can be propagating fields, the free causal propagation of which is described by the usual Feynman propagator~\eqref{eq:photprop}. 
Within the LCFA, we have the additional constraint that the considered momentum transfer
is manifestly restricted to the soft momentum regime, i.e., $\{|p^0|,|\vec{p}|\}\lesssim\upsilon\ll m$ (cf. above).

Here, we argue that this constraint will be fulfilled automatically in
the evaluation of all the Feynman diagrams that can arise as quantum
corrections within the effective theory $\Gamma_\text{eff}$ of
low-frequency photon fields in slowly varying electromagnetic fields.
For this, we first stress that $\Gamma_\text{eff}$ already
incorporates all 1PI proper vertices by definition, such that further
quantum corrections to be evaluated within the effective theory of
low-frequency photon fields must be 1PR.
By construction, the virtual photons in these 1PR diagrams mediate
between slowly varying fields only, since the external lines of the
1PI building blocks are either low-frequency photons or slowly varying
electromagnetic fields.  Hence, the above kinematic constraint is
indeed fulfilled automatically.

In a next step, we utilize $\Gamma_\text{eff}$ to derive some physically relevant explicit results: 
as the prime example, we compute the 1PR contribution to the Heisenberg-Euler effective action $\Gamma^{2\text{-loops}}_{\rm HE}\big|_{1{\rm PR}}$ in slowly varying external fields, introduced and discussed already in Secs.~\ref{sec:HEvs1PI} and \ref{sec:knownHE} above.

\subsection{1PR contribution to the Heisenberg-Euler effective action} \label{subsec:1PR}

Let us now focus on the effective self-interactions of the external electromagnetic field arising in this theory.
At one-loop order these are encoded in $\Gamma^{(0)}_{\rm int}\big|_{1\text{-loop}}=\Gamma_\text{HE}^{1\text{-loop}}$ (cf. Figs.~\ref{fig:1loopGamma} and \ref{fig:S_int}).
At two loops, in addition to
$\Gamma^{(0)}_{\rm int}\big|_{2\text{-loop}}=\Gamma_\text{HE}^{2\text{-loop}}\big|_{1{\rm PI}}$, also the 1PR diagram depicted in Fig.~\ref{fig:2loopGamma} (right) contributes.
It corresponds to the following expression:
\begin{equation}
 \Gamma_\text{HE}^{2\text{-loop}}\big|_{1{\rm PR}}=\frac{1}{2}\int_p j^\mu_{1\text{-loop}}(p) D_{\mu\nu}(p) j^\nu_{1\text{-loop}}(-p) \,, \label{eq:HE_PR}
\end{equation}
where $j^\mu_{l\text{-loop}}:=s_{(1)}^\mu\big|_{l\text{-loop}}$.
We emphasize that the integration in \Eqref{eq:HE_PR}, which is
formally over all virtual momentum transfers, exclusively receives
contributions from the soft momentum regime.  This is because the
photon currents $j^\mu(p)$ only induce low-energy modes by
construction via the LCFA. The constant-field limit in
\Eqref{eq:jgleich0?} provides an obvious example for the underlying
mechanism.
Inserting the explicit expressions for the currents and the photon propagator~\eqref{eq:photprop} in the Feynman gauge, we obtain
\begin{equation}
\Gamma_\text{HE}^{2\text{-loop}}\big|_{1{\rm PR}}
= \frac{1}{2}\int_x\int_{x'} G^{\mu\nu}(x-x')\frac{\partial{\cal L}_{\rm HE}^{1\text{-loop}}}{\partial\bar F^{\mu}_{\ \ \!\alpha}}(x)\frac{\partial{\cal L}_{\rm HE}^{1\text{-loop}}}{\partial\bar F^{\nu\alpha}}(x') \, . \label{eq:DeltaS2_1}
\end{equation}
Here we have defined
\begin{equation}
 G^{\mu\nu}(\tilde x):=4\int_p \frac{p^{\mu}p^{\nu}}{p^2-i\epsilon}{e}^{{i}\tilde xp}
 =
 \frac{2}{\pi^2}\frac{{i}}{(\tilde x^2+i\epsilon)^2}\Bigl(g^{\mu\nu}-4\frac{\tilde x^\mu \tilde x^\nu}{\tilde x^2+i\epsilon}\Bigr) , \label{eq:kint}
\end{equation}
which fulfills $\frac{1}{4}g_{\mu\nu}G^{\mu\nu}(\tilde x)=\delta(\tilde x)$ and $\int_{\tilde x}G^{\mu\nu}(\tilde x)=g^{\mu\nu}$.

Expressing the derivatives for $\bar F$ in terms of derivatives for $\F$ and $\G$ (cf. Appendix~\ref{app:ids}), \Eqref{eq:DeltaS2_1} can be represented as
\begin{multline}
\Gamma_\text{HE}^{2\text{-loop}}\big|_{1{\rm PR}} = \biggl\{\frac{1}{4}\int_x\int_{x'}G_{\alpha\beta}(x-x')\bar F^{\alpha}_{\ \rho}(x) \\
\times\biggl[ \frac{1}{2}\bar F^{\beta\rho}(x')\biggl(\frac{\partial{\cal L}}{\partial{\cal F}}(x)\,\frac{\partial{\cal L}}{\partial{\cal F}}(x')
+ \frac{\partial{\cal L}}{\partial{\cal G}}(x)\,\frac{\partial{\cal L}}{\partial{\cal G}}(x')\biggr) 
 + {}^*\!\bar F^{\beta\rho}(x')\,\frac{\partial{\cal L}}{\partial{\cal F}}(x)\,\frac{\partial{\cal L}}{\partial{\cal G}}(x') \biggr] \\
 - \int_x{\cal F}(x)\Bigl(\frac{\partial{\cal L}}{\partial{\cal G}}(x)\Bigr)^2\biggr\}\bigg|_{{\cal L}={\cal L}_\text{HE}^{1\text{-loop}}} ,
 \label{eq:DeltaS2_2}
\end{multline}
where we employed the identity ${}^*\!\bar F^{\alpha}_{\ \rho}(x)\,{}^*\!\bar F^{\beta\rho}(x') = \bar F^{\beta\rho}(x)\bar F^{\alpha}_{\ \rho}(x')-\frac{1}{2}g^{\alpha\beta}\bar F_{\sigma\rho}(x)\bar F^{\sigma\rho}(x')$.
The products of derivatives of ${\cal L}_\text{HE}^{1\text{-loop}}$ in \Eqref{eq:DeltaS2_2} for $\cal F$ and $\cal G$ can be expressed in terms of double integral representations which follow directly from the parameter integral representation of ${\cal L}_\text{HE}^{1\text{-loop}}$.

Even though derived from a LCFA, \Eqref{eq:DeltaS2_2} gives rise to nonlocal interactions among electromagnetic fields.
However, for slowly varying electromagnetic fields as considered here, these nonlocalities are expected to be very weak.
Particularly for constant external fields, the field strength tensor $\bar F$ and thus the effective Lagrangian become independent of $x$ and $x'$,
such that the integrations over position space in Eqs.~\eqref{eq:DeltaS2_1} and \eqref{eq:DeltaS2_2} can be performed right away, resulting in \Eqref{eq:1PR} above.

Let us finally resolve the seeming
discrepancy that the constant-field limit of Eqs.~\eqref{eq:DeltaS2_1} and \eqref{eq:DeltaS2_2} yields the finite result~\eqref{eq:1PR} even though the formal expression of the photon current vanishes in constant fields; cf. \Eqref{eq:jgleich0?}.
The photon current~\eqref{eq:jgleich0?} vanishes because of
the factor linear in $p^\sigma$ multiplying $\delta(p)$, which ensures the Ward identity $p_\sigma j^\sigma(p)=0$ to hold.
This implies that no real external photons $a_\sigma(p)$, exhibiting a regular behavior for $p^\sigma\to0$, can be induced from $j^\sigma(p)$.
However, this behavior is clearly not fulfilled by virtual photons.
As the photon propagator~\eqref{eq:photprop} scales as $\sim\frac{1}{p^2}$,
the linear momentum dependences of the two individual currents effectively drop out upon combination with two photon currents, leaving us with a finite contribution.

Correspondingly, the two-loop physical effective interaction among generic external electromagnetic fields is determined by the combination
$\Gamma_{\text{HE}}^{\text{2-loop}}=\Gamma_{\text{HE}}^{\text{2-loop}}\big|_{1{\rm PI}}+\Gamma_{\text{HE}}^{\text{2-loop}}\big|_{1{\rm PR}}$.
Of course, similar 1PR diagrams are expected to contribute to the self-interactions of the external electromagnetic field at higher loop orders -- even in constant external fields.
Besides, they obviously also need to be accounted for in determining the effective interactions between any given number of photons in the quantum vacuum subject
to external electromagnetic fields.

\subsection{Low-frequency photon propagation} \label{sec:softphotonprop}

In this section we study quantum corrections to photon propagation -- i.e., photon-photon correlators -- in external electromagnetic fields up to order $(\frac{\alpha}{\pi})^2$.
It is instructive to have a look on the various Feynman diagrams potentially contributing to photon propagation up to this order; see Fig.~\ref{fig:S_int}.
\begin{figure}
 \centering
 \includegraphics[width=0.7\columnwidth]{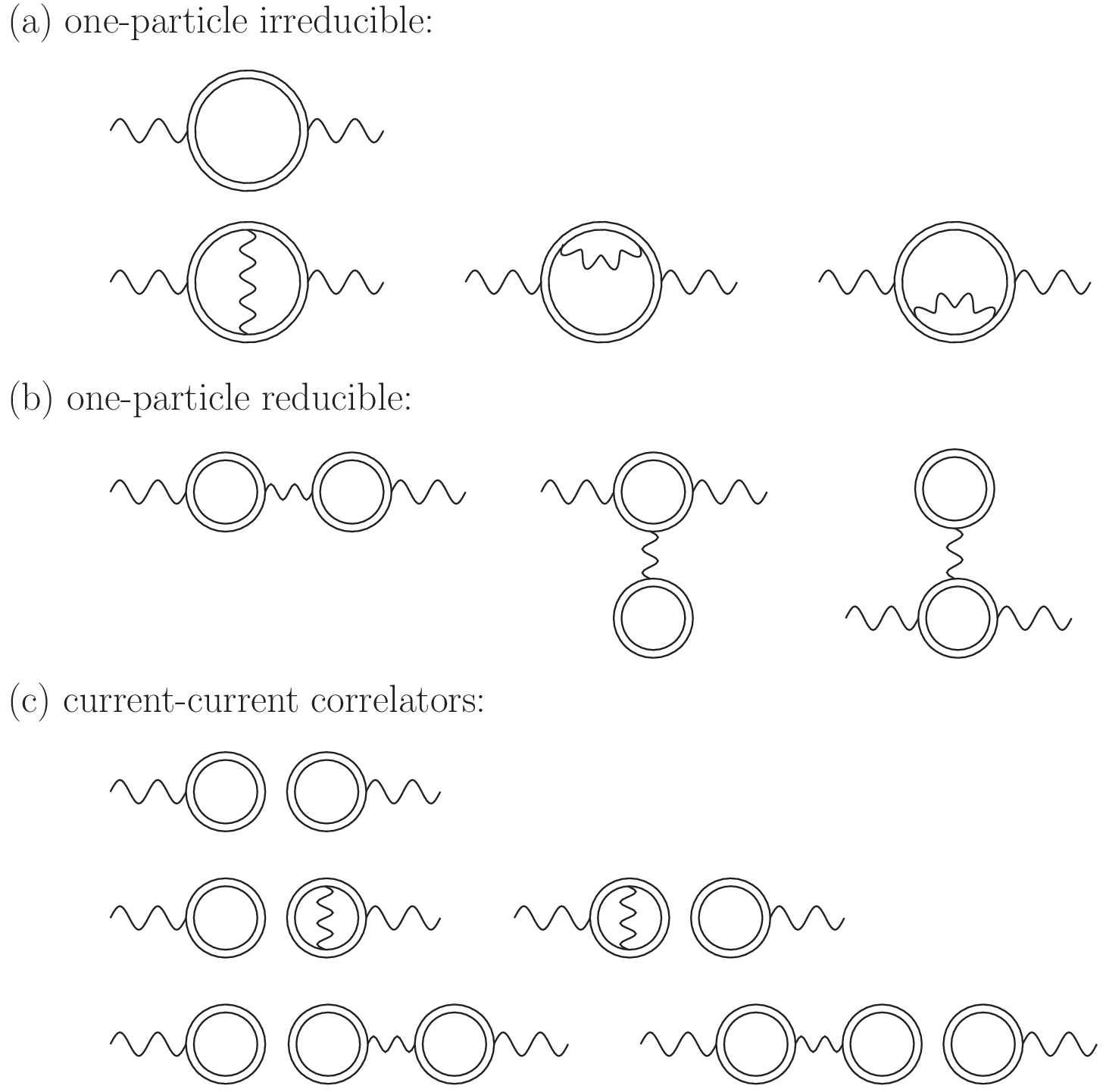}
\caption{Different types of Feynman diagrams contributing to photon propagation in generic external electromagnetic fields at order $\frac{\alpha}{\pi}$ and $(\frac{\alpha}{\pi})^2$; for the definition of the double line, cf. Fig.~\ref{fig:1loopGamma}.
Note that the presence of the current-current diagrams (c) invalidates the equivalence between counting numbers of loops and powers of $\frac{\alpha}{\pi}$.}
\label{fig:pi}
\end{figure}
We organize them into (a) 1PI diagrams, (b) 1PR diagrams, and disconnected contributions which amount to (c) current-current correlators.
Note that there are just two diagrams at order $\frac{\alpha}{\pi}$, namely the first one in Fig.~\ref{fig:pi}(a) and in Fig.~\ref{fig:pi}(c), respectively.
All the other diagrams shown are proportional to $(\frac{\alpha}{\pi})^2$.

The diagrams depicted in Fig.~\ref{fig:pi}(a) constitute the 1PI part of the photon polarization tensor at one (first line) and two loops (second line).
They are contained in $\Gamma_\text{int}^{(2)}$, and are given by
\begin{equation}
 s_{(2)}^{\rho\sigma}(p,p')\big|_{l\text{-loop}}
 = -4 p^\mu p'^\nu 
  \int_x{e}^{{i}(p+p')x}\frac{\partial^2{\cal L}^{l\text{-loop}}_\text{HE}}{\partial\bar F^{\mu}_{\ \ \!\rho}\,\partial\bar F^{\nu\phantom{\beta}}_{\ \ \!\sigma}}(x)\bigg|_{1{\rm PI}} \,; \label{eq:Pi_irred}
\end{equation}
cf. Eqs.~\eqref{eq:S_n}-\eqref{eq:S}. Note that $s_{(2)}^{\rho\sigma}\big|_{1\text{-loop}}$ corresponds to the slowly varying field limit of the one-loop polarization tensor $\Pi^{\rho\sigma}\equiv s_{(2)}^{\rho\sigma}\big|_{1\text{-loop}}$ defined in Sec.~\ref{subsec:diagexp} above.
The definition~\eqref{eq:Pi_irred} automatically accounts for all the topologically inequivalent 1PI diagrams depicted in the second line of Fig.~\ref{fig:pi}(a).
Obviously, we have $s_{(2)}^{\rho\sigma}(p,p')\big|_{l\text{-loop}}\sim(\frac{\alpha}{\pi})^l$.
Expressing the derivatives for $\bar F$ in terms of derivatives for $\F$ and $\G$, \Eqref{eq:Pi_irred} can be represented as \cite{Karbstein:2015cpa}
\begin{multline}
 s_{(2)}^{\rho\sigma}(p,p')\big|_{l\text{-loop}}
 =-\int_x{e}^{{i}(p+p')x} \biggl[
 \bigl((pp')g^{\rho\sigma} - p'^\rho p^\sigma \bigr)\frac{\partial{\cal L}}{\partial{\cal F}}(x)
 + p'_{\mu} p_{\alpha} \epsilon^{\rho\sigma\mu\alpha}\, \frac{\partial{\cal L}}{\partial{\cal G}}(x) \\
 + \bigl(p\bar F(x)\bigr)^\rho  \bigl(p'\bar F(x)\bigr)^\sigma\,\frac{\partial^2 {\cal L}}{\partial{\cal F}^2}(x)
 + \bigl(p{}^*\bar F(x)\bigr)^\rho \bigl(p'\,{}^*\!\bar F(x)\bigr)^\sigma\,\frac{\partial^2 {\cal L}}{\partial{\cal G}^2}(x) \\
 + \bigl[\bigl(p\,{}^*\!\bar F(x)\bigr)^\rho \bigl(p'\bar F(x)\bigr)^\sigma + \bigl(p\bar F(x)\bigr)^\rho \bigl(p'\,{}^*\!\bar F(x)\bigr)^\sigma\,\bigr]\,\frac{\partial^2 {\cal L}}{\partial{\cal F}\partial{\cal G}}(x)
 \biggr]\bigg|_{{\cal L}={\cal L}_\text{HE}^{l\text{-loop}}\big|_{1{\rm PI}}}, \label{eq:Pi}
\end{multline}
where we have employed the shorthand notations $(p\bar F)^\mu=p_\nu\bar F^{\nu\mu}$, $(p\,{}^*\!\bar F)^\mu=p_\nu\,{}^*\!\bar F^{\nu\mu}$, $(pp')=p_{\mu}p'^\mu$, etc.
In the constant-field limit, $\bar F$ as well as ${\cal L}_\text{HE}$ do not depend on the space-time coordinate and \Eqref{eq:Pi} simplifies significantly:
In this limit, the $x$ integration can be performed right away, giving rise to an overall delta function ensuring momentum conservation in constant fields, and thus
\begin{equation}
 s_{(2)}^{\rho\sigma}(p,p')\big|_{l\text{-loop}}=(2\pi)^4\delta(p+p')\sum_{j=0}^3 P_j^{\rho\sigma}(p) \, c_j^{l\text{-loop}}(\F,\G) , \label{eq:Piconst}
\end{equation}
where we have introduced the tensor structures
\begin{align}
 P_0^{\rho\sigma}(p) &= \bigl(p^2 g^{\rho\sigma} - p^\rho p^\sigma \bigr) \,, \nonumber\\
 P_1^{\rho\sigma}(p) &= (p\bar F)^\rho  (p\bar F)^\sigma \,, \nonumber\\
 P_2^{\rho\sigma}(p) &= (p\,{}^*\!\bar F)^\rho (p\,{}^*\!\bar F)^\sigma \,, \nonumber\\
 P_3^{\rho\sigma}(p) &= \bigl[(p\,{}^*\!\bar F)^\rho (p\bar F)^\sigma + (p\bar F)^\rho (p\,{}^*\!\bar F)^\sigma\bigr] \,.  \label{eq:Pmunu}
\end{align}
The associated coefficients $c^{l\text{-loop}}_{j}(\F,\G)$ are given by
\begin{align}
 &c^{l\text{-loop}}_{0}=\frac{\partial{\cal L}_\text{HE}^{l\text{-loop}}}{\partial{\cal F}}\bigg|_{1{\rm PI}}\,, \quad
 c^{l\text{-loop}}_{1}=\frac{\partial^2 {\cal L}_\text{HE}^{l\text{-loop}}}{\partial{\cal F}^2}\bigg|_{1{\rm PI}}\,, \nonumber\\
 &c^{l\text{-loop}}_{2}=\frac{\partial^2 {\cal L}_\text{HE}^{l\text{-loop}}}{\partial{\cal G}^2}\bigg|_{1{\rm PI}}\,, \quad
 c^{l\text{-loop}}_{3}=\frac{\partial^2 {\cal L}_\text{HE}^{l\text{-loop}}}{\partial{\cal F}\partial{\cal G}}\bigg|_{1{\rm PI}}\,.
 \label{eq:cs}
\end{align}
Note that the tensor structure $p'_{\mu} p_{\alpha} \epsilon^{\rho\sigma\mu\alpha}$ vanishes in constant fields, where $p'_{\mu}=-p_{\mu}$.

Let us now have a closer look on the other Feynman diagrams depicted in Fig.~\ref{fig:pi}.
The first diagram in Fig.~\ref{fig:pi}(b) is just an iteration of $\Pi^{\rho\sigma}$, and the corresponding expression reads
\begin{multline}
 \int_k\Pi^{\rho\mu}(p,k) D_{\mu\nu}(k)\Pi^{\nu\sigma}(-k,p') \\
 =4\,p^{\alpha} p'^{\beta}  \int_x\int_{x'}{e}^{{i}(xp+x'p')} G^{\mu\nu}(x-x')
  \frac{\partial^2{\cal L}_\text{HE}^{1\text{-loop}}}{\partial\bar F^{\alpha}_{\ \,\rho}\,\partial\bar F^{\mu}_{\ \ \!\sigma}}(x)
  \frac{\partial^2{\cal L}_\text{HE}^{1\text{-loop}}}{\partial\bar F^{\sigma\nu}\partial\bar F^\beta_{\ \ \sigma}}(x')
   \,. \label{eq:PiDPi}
\end{multline}
A similar diagram exists in the absence of external fields.
Contrarily, all the other diagrams in Fig.~\ref{fig:pi}(b) and those in Fig.~\ref{fig:pi}(c) do not contribute at zero field, because of Furry's theorem.
The last two diagrams in Fig.~\ref{fig:pi}(b) arise from saturating one leg of an effective three-photon coupling with $j_{1\text{-loop}}$.
Both diagrams yield the same result.
Their combined contribution is given by
\begin{multline}
 \int_{k}s_{(3)}^{\rho\sigma\mu}(p,p',-k)\big|_{1\text{-loop}}D_{\mu\nu}(k)j^\nu_{1\text{-loop}}(k) \\
 =4\, p^{\alpha}
 p'^{\beta} \int_x\int_{x'}{e}^{{i}x(p+p')}G^{\mu\nu}(x-x')
 \frac{\partial^3{\cal L}_\text{HE}^{1\text{-loop}}}{\partial\bar F^{\alpha}_{\ \ \rho}\,\partial\bar F^\beta_{\ \ \sigma}\,\partial\bar F^{\mu}_{\ \ \gamma}}(x)
 \frac{\partial{\cal L}_\text{HE}^{1\text{-loop}}}{\partial\bar F^{\gamma\nu}}(x')\,. \label{eq:s3Dj}
\end{multline}
Finally, we turn to the disconnected diagrams in Fig.~\ref{fig:pi}.
The first three diagrams in Fig.~\ref{fig:pi}(c) correspond to $ij_{1\text{-loop}}^{\rho}(p)j_{1\text{-loop}}^{\sigma}(p')$,
$ij_{1\text{-loop}}^{\rho}(p)j_{2\text{-loop}}^{\sigma}(p')$ and $ij_{2\text{-loop}}^{\rho}(p)j_{1\text{-loop}}^{\sigma}(p')$, where
\begin{equation}
 ij_{l\text{-loop}}^{\rho}(p)j_{l'\text{-loop}}^{\sigma}(p')
 = -4i\, p^{\alpha}p'^{\beta} 
 \int_{x}{e}^{{i}xp}\frac{\partial{\cal L}_\text{HE}^{l\text{-loop}}}{\partial\bar F^{\alpha}_{\ \ \!\rho}}(x)\bigg|_{1{\rm PI}}
 \int_{x'}{e}^{{i}x'p'}\frac{\partial{\cal L}_\text{HE}^{l'\text{-loop}}}{\partial\bar F^{\beta\phantom{\beta}}_{\ \ \!\sigma}}(x')\bigg|_{1{\rm PI}} \, . \label{eq:jj}
\end{equation}
The two diagrams depicted in the last line of Fig.~\ref{fig:pi}(c) amount to current-current correlators with one-loop quantum correction to the out- and ingoing photon line, respectively.  
The left one can be expressed as
\begin{multline}
 ij_{l\text{-loop}}^{\rho}(p)\int_k j_{1\text{-loop}}^\mu(k)D_{\mu\nu}(k)\Pi_{1\text{-loop}}^{\nu\sigma}(-k,p') \\
 = 4i\, p^{\alpha} p'^{\beta}
 \int_{x''}{e}^{{i}x''p}\frac{\partial{\cal L}_{1\text{-loop}}}{\partial\bar F^{\alpha}_{\ \ \,\rho}}(x'') 
 \int_{x}\int_{x'}{e}^{{i}x'p'}G^{\mu\nu}(x-x')
 \frac{\partial{\cal L}_\text{HE}^{1\text{-loop}}}{\partial\bar F^{\mu}_{\ \ \!\gamma}}(x)
 \frac{\partial^2{\cal L}_\text{HE}^{1\text{-loop}}}{\partial\bar F^{\gamma\nu}\partial\bar F^{\beta}_{\ \ \,\sigma}}(x') \,, \label{eq:(c)nexttolast}
\end{multline}
and the right one corresponds to \Eqref{eq:(c)nexttolast} with the replacements $p \leftrightarrow p'$ and $\rho\leftrightarrow\sigma$.

In constant electromagnetic fields, all diagrams depicted in Fig.~\ref{fig:pi}(c) vanish if at least one of the external photons is real; cf. the discussion in  Sec.~\ref{subsec:1PR}.
Of course, the derivatives of the Lagrangian for $\bar F$ in Eqs.~\eqref{eq:PiDPi}-\eqref{eq:(c)nexttolast} could again be expressed in terms of derivatives for $\F$ and $\G$ (cf. Appendix~\ref{app:ids}).
While this would allow us to identify the various tensor structures spanning these contributions, the resulting expressions are lengthy so that we do not show them here.

Correspondingly, the 
photon polarization tensor in the presence of an external field is not only given
by 1PI diagrams, but also receives corrections from 1PR and even disconnected diagrams.
More precisely, it is made up of all the diagrams that do not correspond to 
iterations of more elementary diagrams describing quantum corrections to photon propagation.
The full dressed photon propagator in the external field -- accounting for quantum corrections to all orders -- is then obtained by summing up all possible 
iterations analogously to a Dyson series of this photon polarization tensor.
At $l$ loops, we have
\begin{equation}
 \Pi_{l\text{-loop}}^{\rho\sigma}(p,p')=\Pi_{l\text{-loop}}^{\rho\sigma}(p,p')\big|_{1{\rm PI}}+\Delta\Pi_{l\text{-loop}}^{\rho\sigma}(p,p')\,, \label{eq:frakM}
\end{equation}
where $\Pi_{l\text{-loop}}^{\rho\sigma}\big|_{1{\rm PI}}:=s^{\rho\sigma}_{(2)}\big|_{l\text{-loop}}$ and $\Delta\Pi^{\rho\sigma}_\text{2\text{-loop}}$ refers to the contributions of 1PR and disconnected diagrams.
In a slight abuse of nomenclature, we label contributions to the photon polarization tensor which scale as $\sim(\frac{\alpha}{\pi})^l$ with ``$l$-loop'', even though $\Delta\Pi_{l\text{-loop}}^{\rho\sigma}$ generically also includes current-current correlators involving higher loop numbers; cf. Fig.~\ref{fig:pi}.
The explicit expressions for $\Delta\Pi_{l\text{-loop}}^{\rho\sigma}$ at one and two loops are
\begin{equation}
 \Delta\Pi_{1\text{-loop}}^{\rho\sigma}(p,p') := i j_{1\text{-loop}}^{\rho}(p)j_{1\text{-loop}}^{\sigma}(p') \,, \label{eq:DeltaPi1}
\end{equation}
and
\begin{multline}
 \Delta\Pi_{2\text{-loop}}^{\rho\sigma}(p,p') := 
 \int_{k} s_{(3)}^{\rho\sigma\mu}(p,p',-k)\big|_{1\text{-loop}}D_{\mu\nu}(k) j^\nu_{1\text{-loop}}(k) \\
 +i j_{1\text{-loop}}^{\rho}(p)j_{2\text{-loop}}^{\sigma}(p')
 +i j_{2\text{-loop}}^{\rho}(p)j_{1\text{-loop}}^{\sigma}(p')\,. \label{eq:DeltaPi2}
\end{multline}
Equations~\eqref{eq:DeltaPi1} and \eqref{eq:DeltaPi2} account for the five distinct diagrams in Fig.~\ref{fig:pi}(b) and \ref{fig:pi}(c) that do not correspond to 
iterations of more elementary quantum corrections to photon propagation.

In inhomogeneous electromagnetic fields, the explicit expressions for Eqs.~\eqref{eq:DeltaPi1} and \eqref{eq:DeltaPi2} written in terms of derivatives of ${\cal L}_\text{HE}^{1\text{-loop}}$ for the scalar field invariants $\F$ and $\G$ are rather lengthy.
The main reasons for this are the different space-time arguments of the derivatives of ${\cal L}_\text{HE}^{1\text{-loop}}$.
However, in constant external fields these expressions simplify significantly, as all the current-current correlators vanish and the derivatives of ${\cal L}_\text{HE}^{1\text{-loop}}$ with respect to
$\bar F$ become independent of $x$. 
More specifically, in this limit, we obtain $\Delta\Pi_{1\text{-loop}}^{\rho\sigma}(p,p')=0$, and
\begin{align}
 \Delta\Pi_{2\text{-loop}}^{\rho\sigma}(p,p')&= -(2\pi)^4\delta(p+p')4p^{\mu_1}p^{\mu_2}
 \frac{\partial^3{\cal L}_\text{HE}^{1\text{-loop}}}{\partial\bar F^{\mu_1}_{\ \ \,\rho}\partial\bar F^{\mu_2}_{\ \ \,\sigma}\partial\bar F^{\mu\nu_3}}
 \frac{\partial{\cal L}_\text{HE}^{1\text{-loop}}}{\partial\bar F_{\nu_3\mu}} \nonumber\\
 &=(2\pi)^4\delta(p+p')\sum_{j=0}^3 P_j^{\rho\sigma}(p) \, \Delta c^{2\text{-loop}}_j(\F,\G) \,, \label{eq:DeltaPiconst}
\end{align}
with the tensor structures $P_j^{\rho\sigma}(p)$ defined in \Eqref{eq:Pmunu}.
Hence, \Eqref{eq:DeltaPiconst} is spanned by the same tensor structures as $\Pi^{\rho\sigma}\big|_{1{\rm PI}}$ in constant fields as listed in \Eqref{eq:Pmunu}.
The coefficients $\Delta c^{2\text{-loop}}_j$ are given by
\begin{align}
\Delta c^{2\text{-loop}}_0&=\frac{1}{2}(\bar F\partial_{\bar F}{\cal L})\dLFF+\frac{1}{2}(^*\bar F\partial_{\bar F}{\cal L})\dLFG \,, \nonumber\\
 \Delta c^{2\text{-loop}}_1&=\frac{1}{2}(\bar F\partial_{\bar F}{\cal L})\dLFFF + \frac{1}{2}(^*\bar F\partial_{\bar F}{\cal L})\dLFFG + \dLF\dLFF-\dLG\dLFG\,, \nonumber\\
 \Delta c^{2\text{-loop}}_2&=\frac{1}{2}(\bar F\partial_{\bar F}{\cal L})\dLFGG + \frac{1}{2}(^*\bar F\partial_{\bar F}{\cal L})\dLGGG + \dLF\dLGG + \dLG\dLFG\,, \nonumber\\
 \Delta c^{2\text{-loop}}_3&=\frac{1}{2}(\bar F\partial_{\bar F}{\cal L})\dLFFG + \frac{1}{2}(^*\bar F\partial_{\bar F}{\cal L})\dLFGG 
 +\dLF\dLFG +\frac{1}{2}\dLG\Bigl(\dLFF-\dLGG\Bigr)\,, \label{eq:Deltacs}
\end{align}
where ${\cal L}={\cal L}_\text{HE}^{1\text{-loop}}$, and we have made use of the shorthand notations
$ (\bar F\partial_{\bar F}{\cal L}):=\bar F^{\mu\nu}\frac{\partial{\cal L}}{\partial\bar F^{\mu\nu}}=2\bigl(\F \frac{\partial{\cal L}}{\partial{\cal F}}
 + \G \frac{\partial{\cal L}}{\partial{\cal G}}\bigr) $
and
$ (^*\bar F\partial_{\bar F}{\cal L}):={}^*\bar F^{\mu\nu}\frac{\partial{\cal L}}{\partial \bar F^{\mu\nu}}=2\bigl(\G \frac{\partial{\cal L}}{\partial{\cal F}}
 - \F \frac{\partial{\cal L}}{\partial{\cal G}}\bigr) $.

It is instructive to compare the coefficients $\Delta c^{2\text{-loop}}_j$ in \Eqref{eq:Deltacs} with the coefficients $c^{2\text{-loop}}_{j}$ in \Eqref{eq:cs}.
This comparison is rather straightforward in the limits of weak and strong fields.
For spinor QED, the weak-field expressions for the coefficients~\eqref{eq:cs} and \eqref{eq:Deltacs} can be obtained with the help of Eqs.~\eqref{eq:L1loopweakfield} and \eqref{eq:L2loopweakfield}. They are given by
\begin{align}
 c^{2\text{-loop}}_{0} &= \Bigl(\frac{\alpha}{\pi}\Bigr)^2\Bigl[\frac{32}{81}\tilde\F -\frac{1}{45^2}\Bigl(\frac{3657}{2}\,\tilde\F^2+1082\,\tilde\G^2\Bigr)+{\cal O}(\epsilon^6)\Bigr]\,, \nonumber\\
 c^{2\text{-loop}}_{1} &= \frac{e^2}{m^4}\Bigl(\frac{\alpha}{\pi}\Bigr)^2\Bigl[\frac{32}{81} -\frac{3657}{45^2} \tilde\F+{\cal O}(\epsilon^4)\Bigr]\,, \nonumber\\
 c^{2\text{-loop}}_{2} &= \frac{e^2}{m^4}\Bigl(\frac{\alpha}{\pi}\Bigr)^2\Bigl[\frac{263}{324} -\frac{2164}{45^2} \tilde\F+{\cal O}(\epsilon^4)\Bigr]\,, \nonumber\\
 c^{2\text{-loop}}_{3} &= -\frac{e^2}{m^4}\Bigl(\frac{\alpha}{\pi}\Bigr)^2\Bigl[\frac{2164}{45^2} \,\tilde\G+{\cal O}(\epsilon^4)\Bigr] \,,
\end{align}
and
\begin{align}
 \Delta c^{2\text{-loop}}_0 &= \Bigl(\frac{\alpha}{\pi}\Bigr)^2\Bigl[\frac{1}{45^2}(16\,\tilde\F^2+28\,\tilde\G^2)+{\cal O}(\epsilon^6)\Bigr]\,, \nonumber\\
 \Delta c^{2\text{-loop}}_1 &= \frac{e^2}{m^4}\Bigl(\frac{\alpha}{\pi}\Bigr)^2\Bigl[\frac{16}{45^2}\,\tilde\F+{\cal O}(\epsilon^4)\Bigr]\,, \nonumber\\
 \Delta c^{2\text{-loop}}_2 &= \frac{e^2}{m^4}\Bigl(\frac{\alpha}{\pi}\Bigr)^2\Bigl[\frac{28}{45^2}\,\tilde\F+{\cal O}(\epsilon^4)\Bigr]\,, \nonumber\\
 \Delta c^{2\text{-loop}}_3 &= -\frac{e^2}{m^4}\Bigl(\frac{\alpha}{\pi}\Bigr)^2\Bigl[\frac{21}{45^2}\frac{1}{2}\,\tilde\G+{\cal O}(\epsilon^4)\Bigr]\,.
\end{align}
We infer that these coefficients fulfill $\Delta c^{2\text{-loop}}_j/c^{2\text{-loop}}_j={\cal O}(\epsilon^2)$ for $j\in\{0,1,2\}$, and $\Delta c^{2\text{-loop}}_3/c^{2\text{-loop}}_3={\cal O}(1)$.
For $|\tilde\F|\gg1$ and $|\tilde\G|\ll1$, the coefficients $c^{2\text{-loop}}_j$ follow from \Eqref{eq:L2loopstrongfield} by differentiation and read
\begin{align}
 c^{2\text{-loop}}_{0}&=\Bigl(\frac{\alpha}{\pi}\Bigr)^2\Bigl[\frac{1}{4}\ln\sqrt{\tilde\F} +{\cal O}\bigl((\tfrac{1}{\tilde\F})^0\bigr) + {\cal O}(\tilde\G^2)\Bigr] , \nonumber\\
 c^{2\text{-loop}}_{1}&=\frac{1}{\F}\Bigl(\frac{\alpha}{\pi}\Bigr)^2\Bigl[\frac{1}{8} +{\cal O}(1/\sqrt{\tilde\F}) + {\cal O}(\tilde\G^2)\Bigr] , \nonumber\\
 c^{2\text{-loop}}_{2}&=\frac{1}{\F}\Bigl(\frac{\alpha}{\pi}\Bigr)^2\sqrt{{\tilde\F}}\Bigl[-\frac{1}{6\sqrt{2}}  +{\cal O}(1/\sqrt{\tilde\F}) + {\cal O}(\tilde\G^2)\Bigr] , \nonumber\\
 c^{2\text{-loop}}_{3}&=\frac{1}{\F}\Bigl(\frac{\alpha}{\pi}\Bigr)^2 \frac{\tilde\G}{\sqrt{\tilde\F}}\Bigl[\frac{1}{12\sqrt{2}} + {\cal O}(1/\sqrt{\tilde\F}) + {\cal O}(\tilde\G^2)\Bigr] .
\end{align}
The analogous coefficients $\Delta c^{2\text{-loop}}_j$ can be obtained from \Eqref{eq:Deltacs} by using the expressions assembled in \Eqref{eq:1Loopdiffs}, yielding
\begin{align}
 \Delta c^{2\text{-loop}}_0 &=\Bigl(\frac{\alpha}{\pi}\Bigr)^2\Bigl[\frac{1}{18}\ln\sqrt{\tilde\F} +{\cal O}\bigl((\tfrac{1}{\tilde\F})^0\bigr) + {\cal O}(\tilde\G^2)\Bigr] , \nonumber\\
 \Delta c^{2\text{-loop}}_1 &=\frac{1}{\F}\Bigl(\frac{\alpha}{\pi}\Bigr)^2\Bigl[{\cal O}(1/\sqrt{\tilde\F}) + {\cal O}(\tilde\G^2)\Bigr] , \nonumber\\
 \Delta c^{2\text{-loop}}_2 &=\frac{1}{\F}\Bigl(\frac{\alpha}{\pi}\Bigr)^2\sqrt{\tilde\F}\Bigl[\frac{1}{18\sqrt{2}}\ln\sqrt{\tilde\F} +{\cal O}\bigl((\tfrac{1}{\tilde\F})^0\bigr) + {\cal O}(\tilde\G^2)\Bigr] , \nonumber\\
 \Delta c^{2\text{-loop}}_3 &=\frac{1}{\F}\Bigl(\frac{\alpha}{\pi}\Bigr)^2\frac{\tilde\G}{\sqrt{\tilde\F}}\Bigl[-\frac{1}{36\sqrt{2}}\ln\sqrt{\tilde\F} + {\cal O}\bigl((\tfrac{1}{\tilde\F})^0\bigr) + {\cal O}(\tilde\G^2)\Bigr] .
\end{align}
Hence, in the strong-field limit, we read off the scalings $\Delta c^{2\text{-loop}}_0/c^{2\text{-loop}}_0={\cal O}(1)$,
$\Delta c^{2\text{-loop}}_1/c^{2\text{-loop}}_1={\cal O}(1/\sqrt{\tilde\F})$
and $\Delta c^{2\text{-loop}}_2/c^{2\text{-loop}}_2\sim\Delta c^{2\text{-loop}}_3/c^{2\text{-loop}}_3\sim-\frac{1}{3}\ln\sqrt{\tilde\F}$.
This implies that the contribution of $\Delta c^{2\text{-loop}}_1$ is always suppressed in comparison to $c^{2\text{-loop}}_1$ for large values of $|\tilde\F|\gg1$.
By contrast, $\Delta c^{2\text{-loop}}_j$ can surpass $c^{2\text{-loop}}_j$ in magnitude for $j\in\{2,3\}$. Physically, the $c_2$ coefficient is responsible for the enhanced refractive properties of low-frequency photons in a strong field, affecting, e.g., the mode polarized in the plane spanned by a magnetic field and the propagation direction in a magnetized quantum vacuum. 
For completeness, also note that $\Delta c^{2\text{-loop}}_2/c^{1\text{-loop}}_2\sim\Delta c^{2\text{-loop}}_3/c^{1\text{-loop}}_3\sim\frac{1}{6}\frac{\alpha}{\pi}\ln\sqrt{\tilde\F}$; 
cf. the discussion of the analogous considerations for ${\cal L}^{2\text{-loops}}_\text{HE}$ in constant fields in Sec.~\ref{sec:knownHE}.

%%%%%%%%%%%%%%%%%%%%%%%%%%% 
\section{Conclusions and Outlook} \label{sec:Concls+Outl}
%%%%%%%%%%%%%%%%%%%%%%%%%%%

We have taken a fresh look at the famous Heisenberg-Euler effective action $\Gamma_\text{HE}$, which has played a substantial role in the development of quantum field theory, and its relation to the  underlying microscopic theory of QED formulated in terms of the partition function.
We have paid particular attention to
the differences and common ground of $\Gamma_\text{HE}$ and the nowadays more conventional 1PI effective action. 
Most noteworthily and distinctively, $\Gamma_\text{HE}$ also contains 1PR contributions implying quantitative differences to the 1PI effective action from the two-loop level on.

For an efficient determination of these effective actions,
we have constructed an effective theory of low-frequency photons in
the QED vacuum subject to slowly varying electromagnetic fields.
Apart from discussing various generic features of such an effective
theory, our main focus was on the effective interactions generated at
${\cal O}\bigl((\tfrac{\alpha}{\pi})^2\bigr)$.  Here, we in particular
studied the effective self-interaction of external electromagnetic
fields, and derived the photon polarization tensor in the
low-frequency limit. The latter is composed of 1PI, 1PR as well
as disconnected diagrams.

One of our main findings is that the coupling of two one-loop vacuum
currents via a photon propagator gives rise to a nonvanishing 1PR
contribution to $\Gamma_\text{HE}^{2\text{-loop}}$ even in the limit
of constant electromagnetic fields; cf. Fig.~\ref{fig:2loopGamma}
(right).  This contribution was previously believed to vanish.  To
clarify the importance of this newly evaluated 1PR contribution
relatively to the well-known 1PI one, we have investigated
the limits of perturbatively weak and strong fields.
Whereas $\Gamma_\text{HE}^{2\text{-loop}}\big|_{1{\rm PR}}$ is generically
suppressed for weak fields, it can even surpass
$\Gamma_\text{HE}^{2\text{-loop}}\big|_{1{\rm PI}}$ in magnitude for strong fields.
Similar results are obtained for the two-loop photon polarization
tensor. Also here, the 1PR contributions can surpass
the 1PI ones in magnitude for strong fields. 

Our results can also be of relevance beyond QED, for instance, for the exotic case of
a hypothetical minicharged particle sector beyond the Standard Model
of particle physics \cite{Gies:2006ca}.  Beyond QED, the most
essential new feature of the diagram depicted in
Fig.~\ref{fig:2loopGamma} (right) as compared to
Fig.~\ref{fig:2loopGamma} (left) is that the first one can induce
effective interactions mediated by two different fermion species (the
two loops in this diagram do not necessarily have to contain the same
fermion species), while the latter one features a single fermion loop
and thus may only involve one fermion species.
This implies a parametrically different dependence of electromagnetic or optical
observables on the various coupling and mass parameters starting at two-loop level.
As the considerations invoked here can also be adopted for scalar QED,
particularly in a combination of scalar and spinor QED, this type of
mixed effective interactions can also be generated when one of
the loops traces over fermions and the other one over bosons.  For 1PI
diagrams, such an effective coupling of different species can only
happen at three loops or beyond.

%%%%%%%%%%%%%%%%%%%%%%%%%%%%
\section*{Note added}
%%%%%%%%%%%%%%%%%%%%%%%%%%%%

In the present version (v4), we have corrected Eqs. (32), (35), (39), (47), (48), (50) and (B7) by a factor of $1/2$ that was missing in the previous and the published versions.

%%%%%%%%%%%%%%%%%%%%%%%%%%%%
\section*{Acknowledgments}
%%%%%%%%%%%%%%%%%%%%%%%%%%%%

We acknowledge support by the BMBF under grant No. 05P15SJFAA
(FAIR-APPA-SPARC). The authors wish to thank W.~Dittrich for helpful
conversations and for carefully reading the manuscript.
We are particularly grateful to S.~Evans for correspondence that lead to the 
improved version (v4) of this paper.

\appendix

\section{Useful derivative identities} \label{app:ids}

The Heisenberg-Euler effective Lagrangian can be expressed as  ${\cal L}_\text{HE}=\sum_{l=0}^\infty{\cal L}_\text{HE}^{l\text{-loop}}$, where ${\cal L}_\text{HE}^{0\text{-loop}}=-{\cal F}$ is the Maxwell Lagrangian of classical electrodynamics, and
${\cal L}_\text{HE}^{l\text{-loop}}$ with $l\geq1$ encodes quantum corrections vanishing in the formal limit $\hbar\to0$.
In constant fields, we have ${\cal L}_\text{HE}^{l\text{-loop}}\equiv{\cal L}_\text{HE}^{l\text{-loop}}({\cal F},{\cal G}^2)$; cf. the main text for more details.
As ${\cal L}_\text{HE}^{l\text{-loop}}$ is a function of the scalar invariants of the electromagnetic field $\cal F$ and ${\cal G}^2$ only, it is convenient to trade derivatives with respect to the field strength tensor $\bar F$ for derivatives with respect to $\cal F$ and ${\cal G}$.
The explicit expressions for these transformations up to cubic order in the derivative for $\bar F$  are
\begin{equation}
 \frac{\partial{\cal L}_\text{HE}^{l\text{-loop}}}{\partial \bar F^{\mu\nu}}=\frac{1}{2}\biggl(\bar F_{\mu\nu} \frac{\partial}{\partial{\cal F}}  + {}^*\bar F_{\mu\nu} \frac{\partial}{\partial{\cal G}}\biggr){\cal L}_\text{HE}^{l\text{-loop}} \,,
 \label{eq:dL/dF}
\end{equation}
\begin{multline}
 \frac{\partial^2{\cal L}_\text{HE}^{l\text{-loop}}}{\partial \bar F^{\alpha\beta}\partial \bar F^{\mu\nu}}
 =\frac{1}{4}\biggl[\bigl(g_{\alpha\mu}g_{\beta\nu}-g_{\alpha\nu}g_{\beta\mu}\bigr)\frac{\partial}{\partial{\cal F}}  +  \epsilon_{\mu\nu\alpha\beta}\, \frac{\partial}{\partial{\cal G}} 
 + \bar F_{\alpha\beta} \bar F_{\mu\nu}\frac{\partial^2}{\partial{\cal F}^2} + {}^*\bar F_{\alpha\beta}{}^*\bar F_{\mu\nu}\frac{\partial^2}{\partial{\cal G}^2} \\
 + \bigl({}^*\bar F_{\alpha\beta}\bar F_{\mu\nu} + \bar F_{\alpha\beta}{}^*\bar F_{\mu\nu}\bigr)\frac{\partial^2}{\partial{\cal F}\partial{\cal G}}\biggr]{\cal L}_\text{HE}^{l\text{-loop}} \,,
 \label{eq:ddL/dFdF}
\end{multline}
and
\begin{multline}
 \frac{\partial^3{\cal L}_\text{HE}^{l\text{-loop}}}{\partial \bar F^{\rho\sigma}\partial \bar F^{\alpha\beta}\partial \bar F^{\mu\nu}}
 =\frac{1}{8}\biggl\{
   \bar F_{\rho\sigma}\bar F_{\alpha\beta}\bar F_{\mu\nu} \frac{\partial^3}{\partial{\cal F}^3} + {}^*\bar F_{\rho\sigma}{}^*\bar F_{\alpha\beta}{}^*\bar F_{\mu\nu} \frac{\partial^3}{\partial{\cal G}^3} \\
 + \bigl(\bar F_{\rho\sigma}{}^*\bar F_{\alpha\beta}\bar F_{\mu\nu} + \bar F_{\rho\sigma}\bar F_{\alpha\beta}{}^*\bar F_{\mu\nu}+ {}^*\bar F_{\rho\sigma}\bar F_{\alpha\beta}\bar F_{\mu\nu}\bigr)\frac{\partial^3}{\partial{\cal F}^2\partial{\cal G}} \\
 + \bigl({}^*\bar F_{\rho\sigma}{}^*\bar F_{\alpha\beta}\bar F_{\mu\nu} + {}^*\bar F_{\rho\sigma}\bar F_{\alpha\beta}{}^*\bar F_{\mu\nu} + \bar F_{\rho\sigma}{}^*\bar F_{\alpha\beta}{}^*\bar F_{\mu\nu}\bigr) \frac{\partial^3}{\partial{\cal F}\partial{\cal G}^2}\\
 + \Bigl[\bigl(g_{\alpha\mu}g_{\beta\nu}-g_{\alpha\nu}g_{\beta\mu}\bigr)\bar F_{\rho\sigma}
 + \bigl(g_{\rho\mu}g_{\sigma\nu}-g_{\rho\nu}g_{\sigma\mu}\bigr)\bar F_{\alpha\beta}
 + \bigl(g_{\rho\alpha}g_{\sigma\beta}-g_{\rho\beta}g_{\sigma\alpha}\bigr)\bar F_{\mu\nu} \Bigr]\frac{\partial^2}{\partial{\cal F}^2}\\
 + \bigl(\epsilon_{\mu\nu\alpha\beta}\, {}^*\bar F_{\rho\sigma} + \epsilon_{\mu\nu\rho\sigma}{}^*\bar F_{\alpha\beta} + \epsilon_{\alpha\beta\rho\sigma}{}^*\bar F_{\mu\nu} \bigr)\frac{\partial^2}{\partial{\cal G}^2}\\
 + \Bigl[\bigl(\epsilon_{\mu\nu\alpha\beta} \bar F_{\rho\sigma}
 + \epsilon_{\mu\nu\rho\sigma}\bar F_{\alpha\beta}
 + \epsilon_{\alpha\beta\rho\sigma}\bar F_{\mu\nu}
 \bigr)
 +\bigl(g_{\alpha\mu}g_{\beta\nu}-g_{\alpha\nu}g_{\beta\mu}\bigr){}^*\bar F_{\rho\sigma} \\
 + (g_{\rho\mu}g_{\sigma\nu}-g_{\rho\nu}g_{\sigma\mu}){}^*\bar F_{\alpha\beta}
 + (g_{\rho\alpha}g_{\sigma\beta}-g_{\rho\beta}g_{\sigma\alpha}){}^*\bar F_{\mu\nu}
 \Bigr] \frac{\partial^2}{\partial{\cal F}\partial{\cal G}}\biggr\}{\cal L}_\text{HE}^{l\text{-loop}} \,.
 \label{eq:dddL/dFdFdF}
\end{multline}

\section{Strong-field asymptotics for spinor QED} \label{app:sfa}

In this appendix, we concentrate on constant fields and focus on the limit of $|\tilde{\cal F}|=\frac{1}{2}\bigl|(\frac{e\vec{B}}{m^2})^2-(\frac{e\vec{E}}{m^2})^2\bigr|\gg1$ and $|\tilde{\cal G}|=\bigl|\frac{e\vec{E}}{m^2}\cdot\frac{e\vec{B}}{m^2}\bigr|\ll1$, which is of relevance for either strong electric or magnetic fields.

\subsection{One-loop Heisenberg-Euler effective Lagrangian in constant external fields}\label{subsec:1loop}

For the special case of $\G=0$, the one-loop Heisenberg-Euler effective Lagrangian has the following closed-form representation \cite{Dittrich:1975au,Dittrich:1985yb,Dunne:2004nc},
\begin{equation}
 \frac{{\cal L}_\text{HE}^{1\text{-loop}}}{m^4}\bigg|_{\G=0}=\frac{1}{4\pi^2}\frac{1}{2\chi^2}\biggl\{\zeta'(-1,\chi)+\frac{\chi}{2}\Bigl[\bigl(1-\chi\bigr)\ln\chi+\frac{\chi}{2}\Bigr]-\frac{1}{12}\bigl(\ln\chi+1\bigr)\biggr\} , \label{eq:L1loopG=0}
\end{equation}
where $\chi = \frac{1}{2}/\sqrt{2\tilde{\cal F}}$ and $\zeta'(s,\chi)=\partial_s\zeta(s,\chi)$ denotes the first derivative of the Hurwitz zeta function; cf. \Eqref{eq:tF} for the definition of the square root of $\tilde\F$.
In principle, similar closed-form expressions can be obtained for any derivative $\frac{\partial^{n_1+n_2}{\cal L}_\text{HE}^{1\text{-loop}}}{\partial\F^{n_1}\partial\G^{n_2}}\big|_{\G=0}$ with $\{n_1,n_2\}\in\mathbb{N}_0$.
Here we only provide the following explicit expressions \cite{Karbstein:2015cpa},
\begin{align}
 \frac{\partial{\cal L}_\text{HE}^{1\text{-loop}}}{\partial{\cal F}}\bigg|_{{\cal G}=0}
 &=\frac{\alpha}{\pi}\biggl\{4\zeta'(-1,\chi)-\chi\bigl[2\zeta'(0,\chi)-\ln\chi+\chi\bigr]-\frac{1}{3}\ln\chi-\frac{1}{6}\biggr\},
 \nonumber\\
 \frac{\partial^2{\cal L}_\text{HE}^{1\text{-loop}}}{\partial{\cal G}^2}\bigg|_{{\cal G}=0}
 &= \frac{1}{2{\cal F}}\frac{\alpha}{\pi}\biggl\{4\zeta'(-1,\chi)-\chi\bigl[2\zeta'(0,\chi)-\ln\chi+\chi\bigr]-\frac{1}{6}\biggl[2\psi(\chi)+\frac{1}{\chi}+1\biggr]\biggr\}, \label{eq:dL1loopdG2G=0}
\end{align}
where $\psi(\chi)=\frac{\rm d}{{\rm d}\chi}\ln\Gamma(\chi)$ is the Digamma function.
Obviously, we can write 
\begin{equation}
 \frac{{\cal L}_\text{HE}^{1\text{-loop}}}{m^4}=\frac{{\cal L}_\text{HE}^{1\text{-loop}}}{m^4}\bigg|_{\G=0}+\frac{1}{2m^4}\frac{\partial^2{\cal L}_\text{HE}^{1\text{-loop}}}{\partial{\cal G}^2}\bigg|_{\G=0}\,\G^2+{\cal O}(\G^4)\, \label{eq:L1loopsf}
\end{equation}
such that, upon insertion of Eqs.~\eqref{eq:L1loopG=0} and \eqref{eq:dL1loopdG2G=0}, \Eqref{eq:L1loopsf} provides us with a closed-form expression of ${\cal L}_\text{HE}^{1\text{-loop}}/m^4$ in the limit of $|\tilde\G|\ll1$.
Aiming at extracting the asymptotics of \Eqref{eq:L1loopsf} for $|\tilde{\cal F}|\gg1\,\leftrightarrow\,|\chi|\ll1$, it is convenient to employ \cite{dlmf:1}
\begin{equation}
 \zeta'(0,\chi)=\ln\Gamma(\chi)-\frac{1}{2}\ln(2\pi)\,,
\end{equation}
and Eq.~(1.50) of \cite{Dunne:2004nc} (cf. also Appendix~D.6 of \cite{Dittrich:2000zu}),
\begin{equation}
 \zeta'(-1,\chi)=\int_0^\chi{\rm d t \ln\Gamma(t)}+\frac{\chi}{2}(\chi-1)-\frac{\chi}{2}\ln(2\pi)+\zeta'(-1)\,, \label{eq:zetas-1}
\end{equation}
where $\Gamma(.)$ is the Gamma function, and  $\zeta'(-1)\approx -0.16542114$ is the first derivative of the Riemann zeta function evaluated at $-1$.
The leading terms of an expansion of $\ln\Gamma(.)$ for small arguments can then be inferred from Eq.~(1.53) \cite{Dunne:2004nc}.
As a result, we obtain \Eqref{eq:L1loopstrongfield} in the main text.
The asymptotics of the various derivatives of ${\cal L}_\text{HE}^{1\text{-loop}}$ for $\F$ and $\G$ determining ${\cal L}_\text{HE}^{2\text{-loop}}\big|_{1{\rm PR}}$, $c^{(n)}_{1\text{-loop}}$ and $\Delta c^{(n)}_{2\text{-loop}}$ can be inferred along the same lines and read
\begin{align}
 \frac{\partial{\cal L}_\text{HE}^{1\text{-loop}}}{\partial{\cal F}}
 &=\frac{\alpha}{\pi}\biggl[\frac{1}{3}\ln\sqrt{\tilde{\cal F}}+{\cal O}\bigl((\tfrac{1}{\tilde\F})^0\bigr)+{\cal O}(\tilde\G^2)\biggr] , \nonumber\\
 \frac{\partial{\cal L}_\text{HE}^{1\text{-loop}}}{\partial{\cal G}}
 &=\frac{\alpha}{\pi}\frac{\tilde\G}{\sqrt{\tilde\F}}\biggl[\frac{1}{3\sqrt{2}} +{\cal O}(1/\sqrt{\tilde\F}) + {\cal O}(\tilde\G^2)\biggr] , \nonumber\\
 \frac{\partial^2{\cal L}_\text{HE}^{1\text{-loop}}}{\partial{\cal F}^2}
 &=\frac{1}{{\cal F}}\frac{\alpha}{\pi}\biggl[\frac{1}{6}+{\cal O}(1/\sqrt{\tilde\F})+{\cal O}(\tilde\G^2)\biggr] , \nonumber\\
 \frac{\partial^2{\cal L}_\text{HE}^{1\text{-loop}}}{\partial{\cal G}^2}
 &= \frac{1}{{\cal F}}\frac{\alpha}{\pi}\sqrt{\tilde{\cal F}}\biggl[\frac{1}{3\sqrt{2}} +{\cal O}(1/\sqrt{\tilde\F})+{\cal O}(\tilde\G^2)\biggr] , \nonumber\\
 \frac{\partial^2{\cal L}_\text{HE}^{1\text{-loop}}}{\partial{\cal F}\partial{\cal G}}
 &=\frac{1}{\F}\frac{\alpha}{\pi}\frac{\tilde\G}{\sqrt{\tilde\F}}\biggl[-\frac{1}{6\sqrt{2}} +{\cal O}(1/\sqrt{\tilde\F}) + {\cal O}(\tilde\G^2)\biggr] , \nonumber\\
 \frac{\partial^3{\cal L}_\text{HE}^{1\text{-loop}}}{\partial{\cal F}^3}
 &=\frac{1}{{\cal F}^2}\frac{\alpha}{\pi}\biggl[-\frac{1}{6}+{\cal O}(1/\sqrt{\tilde\F})+{\cal O}(\tilde\G^2)\biggr] , \nonumber\\
 \frac{\partial^3{\cal L}_\text{HE}^{1\text{-loop}}}{\partial\F\partial{\cal G}^2}
 &=\frac{1}{{\cal F}^2}\frac{\alpha}{\pi}\sqrt{\tilde{\cal F}}\biggl[-\frac{1}{6\sqrt{2}}+{\cal O}(1/\sqrt{\tilde\F})+{\cal O}(\tilde\G^2)\biggr] , \nonumber\\
 \frac{\partial^3{\cal L}_\text{HE}^{1\text{-loop}}}{\partial{\cal F}^2\partial{\cal G}}
 &=\frac{1}{\F^2}\frac{\alpha}{\pi}\frac{\tilde\G}{\sqrt{\tilde\F}}\biggl[\frac{1}{4\sqrt{2}} +{\cal O}(1/\sqrt{\tilde\F}) + {\cal O}(\tilde\G^2)\biggr] , \nonumber\\
 \frac{\partial^3{\cal L}_\text{HE}^{1\text{-loop}}}{\partial^3{\cal G}}
 &=\frac{1}{\F^2}\frac{\alpha}{\pi}\frac{\tilde\G}{\sqrt{\tilde\F}}\biggl[{\cal O}\bigl((\tfrac{1}{\tilde\F})^0\bigr) + {\cal O}(\tilde\G^2)\biggr] . \label{eq:1Loopdiffs}
\end{align}

\subsection{Two-loop effective Lagrangian in constant external fields}

As detailed in the main text, the two-loop Heisenberg-Euler effective Lagrangian consists of a one-particle irreducible and a one-particle reducible contribution, ${\cal L}_\text{HE}^{2\text{-loop}}={\cal L}_\text{HE}^{2\text{-loop}}\big|_\text{1PI}+{\cal L}_\text{HE}^{2\text{-loop}}\big|_\text{1PR}$; cf. also Fig.~\ref{fig:2loopGamma}.
The one-particle reducible contribution ${\cal L}_\text{HE}^{2\text{-loop}}\big|_\text{1PR}$ follows straightforwardly from ${\cal L}_\text{HE}^{1\text{-loop}}$ via \Eqref{eq:1PR}.
For the special case of ${\cal G}=0$, it has the following closed-form representation:
\begin{align}
\frac{{\cal L}_\text{HE}^{2\text{-loop}}\big|_\text{1PR}}{m^4}\bigg|_{\G=0}
 &=  \frac{\alpha}{\pi}\frac{1}{4\pi^2}\frac{1}{4\chi^2} \biggl\{2\zeta'(-1,\chi)-\frac{\chi}{2}\bigl[2\zeta'(0,\chi)-\ln\chi+\chi\bigr]-\frac{1}{6}\Bigl(\ln\chi+\frac{1}{2}\Bigr)\biggr\}^2  ; \label{eq:DeltaS2b}
\end{align} 
cf. also Appendix~\ref{subsec:1loop}.

Closed-form expressions exists for ${\cal L}_\text{HE}^{2\text{-loop}}\big|_{1{\rm PI}}$ only for the simplified case of self-dual fields \cite{Dunne:2001pp}, but not for the cases of interest here, not even for the special case of ${\G=0}$.
However, the leading strong-field behavior of ${\cal L}_\text{HE}^{2\text{-loop}}\big|_{1{\rm PI}}/m^4$ for $|\tilde\F|\gg1$ and $|\tilde\G|\ll1$ is known explicitly \cite{Ritus:1975} (cf. also \cite{Dittrich:1985yb} for an independent verification).
For $\tilde\F>0$, it is given by
\begin{equation}
 \frac{{\cal L}_\text{HE}^{2\text{-loop}}\big|_{1{\rm PI}}}{m^4}\sim\frac{\alpha}{\pi}\frac{1}{4\pi^2}\frac{\tilde\eta^2}{8}\Bigl(\ln\tilde\eta+\text{constant}\Bigr)+\ldots \,,
\end{equation}
where $\tilde\eta$ is one of the secular invariants of the electromagnetic field, defined as 
\begin{equation}
 \tilde\eta=(\sqrt{\tilde{\cal F}^2+\tilde{\cal G}^2}+\tilde{\cal F})^{1/2} \quad\text{and}\quad \tilde\epsilon=(\sqrt{\tilde{\cal F}^2+\tilde{\cal G}^2}-\tilde{\cal F})^{1/2} \,. \label{eq:strongfield2loopRitus}
\end{equation}
The analogous expression for $\tilde\F<0$ follows from \Eqref{eq:strongfield2loopRitus} by the transformation $\tilde\eta\leftrightarrow-{i}\tilde\epsilon$.
Note that for $\tilde\F\gg1$, we have
\begin{equation}
 \tilde\eta=\sqrt{2\tilde\F}\Bigl(1+{\cal O}\bigl((\tfrac{\tilde\G^2}{\tilde\F^2})^2\bigr)\Bigr) \quad\text{and}\quad \tilde\epsilon=\frac{|\tilde\G|}{\sqrt{2\tilde\F}}\Bigl(1+{\cal O}\bigl((\tfrac{\tilde\G^2}{\tilde\F^2})^2\bigr)\Bigr) \,. \label{eq:lims}
\end{equation}

Aiming at determining the leading strong-field asymptotics of $\Pi^{\mu\nu}_{2\text{-loops}}(p,p')\big|_\text{const.}$, we need the complete scaling of the leading contribution $\sim\tilde\G^2$, for which the terms given in \Eqref{eq:strongfield2loopRitus} are not sufficient.

For this, we have to consider the exact double integral representation of ${\cal L}_\text{HE}^{2\text{-loop}}\big|_{1{\rm PI}}$ in constant external fields \cite{Ritus:1975}.
In the notation of \cite{Ritus:1975}, but adopting the sequential substitutions $s'\to s\nu$, $e\eta s\to-{i}\tau$, introducing
the dimensionless parameters $\tilde\eta=\frac{e\eta}{m^2}$, $\tilde\epsilon=\frac{e\epsilon}{m^2}$ and defining $\kappa\equiv\frac{\tilde\epsilon}{\tilde\eta}$, ${\cal L}_\text{HE}^{2\text{-loop}}\big|_{1{\rm PI}}$ is given by
\begin{equation}
 \frac{{\cal L}_\text{HE}^{2\text{-loop}}\big|_{1{\rm PI}}}{m^4}=\frac{\alpha}{\pi}\frac{\tilde\eta^2}{16\pi^2}\int_0^\infty{\rm d}\tau \int_0^1{\rm d}\nu\,\biggl\{K(\tau,\nu)-\frac{K_0(\tau)}{\nu}
 +  K_0(\tau)\biggl[\ln\Bigl(\frac{\tau}{\tilde\eta}\Bigr)+\gamma-\frac{5}{6}\biggr]\biggr\} \,, \label{eq:L2}
\end{equation}
with
\begin{equation}
 K_0(\tau)={e}^{-\frac{\tau}{\tilde\eta}}\Bigl(\frac{4}{\tilde\eta}-\partial_\tau\Bigr)\biggl[\frac{\kappa}{\tanh(\tau)\tan(\kappa \tau)}-\frac{1}{\tau^2}-\frac{1-\kappa^2}{3}\biggr],
\end{equation}
and
\begin{multline}
 K(\tau,\nu)={e}^{-\frac{\tau}{\tilde\eta}(1+\nu)}\biggl\{\frac{\kappa^2}{PP'}\biggl[\frac{4}{\tilde\eta}(SS'+PP')I_0+2I\biggr]\tau \\
 -\frac{1}{\nu(1+\nu)\tau^3}\biggl[\frac{4}{\tilde\eta}\tau+\frac{2}{1+\nu}-\frac{1-\kappa^2}{3}\tau^2 \biggl(\frac{2}{\tilde\eta}\bigl(\nu -2-2\nu^2 \bigr)\tau
 +\frac{5\nu }{1+\nu}\biggr)\biggr]\biggr\} \, . \label{eq:K}
\end{multline}
Here, we have used
\begin{gather}
 S(\tau)=\cosh(\tau)\cos(\kappa\tau) \,, \quad P(\tau)=\sinh(\tau)\sin(\kappa\tau) \,, \nonumber\\
 S'=S(\nu\tau) \,, \quad P'=P(\nu\tau) \,, \nonumber\\
 I_0=\frac{1}{(b-a)}\ln\Bigl(\frac{b}{a}\Bigr) \,, \quad I=\frac{(q-p)}{(b-a)^2}\ln\Bigl(\frac{b}{a}\Bigr)-\frac{1}{(b-a)}\Bigl(\frac{q}{b}-\frac{p}{a}\Bigr) \,, \nonumber\\
 a=\coth(\tau)+\coth(\nu\tau) \,, \quad b=\kappa\bigl[\cot(\kappa\tau)+\cot(\kappa\nu\tau)\bigr] \,, \nonumber\\
 p=\frac{\cos[\kappa(1-\nu)\tau]}{\sinh(\tau)\sinh(\nu\tau)} \,, \quad q=\frac{\kappa^2\cosh[(1-\nu)\tau]}{\sin(\kappa\tau)\sin(\kappa\nu\tau)} \,. 
\end{gather}

Without loss of generality, we subsequently focus on the limit of $\frac{1}{\tilde\eta}\to0$; the opposite limit of $\frac{1}{\tilde\epsilon}\to0$ can be easily obtained by the transformation $\tilde\eta\leftrightarrow-{i}\tilde\epsilon$.
Due to the overall exponential suppression of both $K_0$ and $K$ with ${e}^{-\frac{\tau}{\tilde\eta}}$, the dominant contributions to ${\cal L}_\text{HE}^{2\text{-loop}}\big|_{1{\rm PI}}$ in the limit of $\frac{1}{\tilde\eta}\to0$ stem from large values of $\tau$.
In a first step we infer that
\begin{equation}
 K_0(\tau)={e}^{-\frac{\tau}{\tilde\eta}}\biggl\{\partial_\tau\biggl[\frac{1}{\tau^2}-\frac{\kappa}{\tanh(\tau)\tan(\kappa \tau)}+\frac{1-\kappa^2}{3}\biggr]+{\cal O}\bigl((\tfrac{1}{\tilde\eta})^1\bigr)\biggr\} \label{eq:K0lim}
\end{equation}
and
\begin{equation}
 K(\tau,\nu)={e}^{-\frac{\tau}{\tilde\eta}(1+\nu)}\biggl\{\frac{\kappa^2}{PP'}2I\tau
 -\frac{1}{(1+\nu)^2}\biggl[\frac{2}{\nu}\frac{1}{\tau^2}-5\frac{1-\kappa^2}{3}\biggr]\frac{1}{\tau}+{\cal O}\bigl((\tfrac{1}{\tilde\eta})^1\bigr)\biggr\} \,. \label{eq:Klim}
\end{equation}
We have explicitly checked that the terms denoted by ${\cal O}\bigl((\tfrac{1}{\tilde\eta})^1\bigr)$ in Eqs.~\eqref{eq:K0lim} and \eqref{eq:Klim} do not increase with $\tau$ for $\tau\to\infty$, but scale at least as ${\cal O}\bigl((\frac{1}{\tau})^0\bigr)$ at any given order in an expansion in $\kappa\to0$. 
Moreover, note that at any fixed order in $\kappa\to0$, we have $\lim_{\tau\to\infty}\bigl|\frac{I}{PP'}\tau\bigr|\sim\tau^l{e}^{-\tau(1+2\nu)}\to0$, with $l\in\mathbb{Z}_0$.
Herewith, we obtain
\begin{equation}
 \int_0^\infty{\rm d}\tau\,K_0(\tau)
 ={\cal O}\bigl((\tfrac{1}{\tilde\eta})^0\bigr){\cal O}(\kappa^0),
\end{equation}
and
\begin{equation}
 \int_0^\infty{\rm d}\tau\,K(\tau,\nu)=\frac{5}{(1+\nu)^2}\frac{1-\kappa^2}{3}\ln\tilde\eta+{\cal O}\bigl((\tfrac{1}{\tilde\eta})^0\bigr){\cal O}(\kappa^0)\,.
\end{equation}
Moreover, we are interested in the following integral:
\begin{multline}
 \int_{0}^\infty{\rm d}\tau\, K_0(\tau)\biggl[\ln(\tfrac{\tau}{\tilde\eta})+\gamma-\frac{5}{6}\biggr]
=\int_{0}^\infty\frac{{\rm d}\tau}{\tau}\,{e}^{-\frac{\tau}{\tilde\eta}}\biggl[\frac{\coth(\tau)}{\tau}-\frac{1}{\tau^2}-\frac{1}{3}\biggr] \\
 +\frac{\kappa^2}{3}\int_{0}^\infty{\rm d}\tau\,{e}^{-\frac{\tau}{\tilde\eta}}\biggl[\frac{1}{\tau}-\coth(\tau)\biggr] + {\cal O}\bigl((\tfrac{1}{\tilde\eta})^0\bigr){\cal O}(\kappa^0) + {\cal O}(\kappa^4)\,. \label{eq:Int1}
\end{multline}
In order to arrive at this result we have made use of \Eqref{eq:K0lim} for $K_0(\tau)$ and performed an integration by parts. 
Thereafter, we have employed an expansion in $\kappa\to0$, keeping terms up to order $\kappa^2$ only.
The integrals in \Eqref{eq:Int1} can be carried out with formulae 3.381.4 and 3.551.3 of \cite{Gradshteyn}:
$\int_0^\infty\frac{{\rm d}\tau}{\tau}\, \tau^{\nu}\, {e}^{-\beta \tau} = \beta^{-\nu}\,\Gamma(\nu)$
and $\int_0^\infty\frac{{\rm d}\tau}{\tau}\, \tau^{\nu}\, {e}^{-\beta \tau} \coth(\tau) = \bigl[2^{1-\nu}\zeta(\nu,\tfrac{\beta}{2})-\beta^{-\nu}\bigr]\Gamma(\nu)$,
valid for $\Re(\beta)>0$ and under
certain conditions on $\nu$, which are rendered irrelevant upon combination of these formulae in performing the manifestly finite integrals in \Eqref{eq:Int1}.
Such integral expressions are common in strong-field QED; cf., e.g., \cite{Tsai:1975iz,Dittrich:1975au,Dittrich:1985yb,Dittrich:2000zu,Karbstein:2013ufa,Karbstein:2015cpa}.
We infer
\begin{equation}
 \int_0^\infty{\rm d}\tau\,{e}^{-\beta\tau}\biggl[\frac{1}{\tau}-\coth(\tau)\biggr]=\psi\Bigl(\frac{\beta}{2}\Bigr)+\frac{1}{\beta}-\ln\frac{\beta}{2}
 =-\frac{1}{\beta}-\ln\beta +{\cal O}(\beta^0)\,, \label{eq:I1}
\end{equation}
where $\psi(\chi)$ is the Digamma function (cf. Sec.~\ref{subsec:1loop}), and
\begin{multline}
 \int_{0}^\infty\frac{{\rm d}\tau}{\tau}\,{e}^{-\beta\tau}\biggl[\frac{\coth(\tau)}{\tau}-\frac{1}{\tau^2}-\frac{1}{3}\biggr] \\
 =\frac{1}{3}\Bigl(\ln\frac{\beta}{2}+1\Bigr)-\beta\ln\frac{\beta}{2}-4\zeta'(-1,\tfrac{\beta}{2})+\frac{\beta^2}{2}\Bigl(\ln\frac{\beta}{2}-\frac{1}{2}\Bigr)
 =\frac{1}{3}\ln\beta + {\cal O}(\beta^0)\,, \label{eq:I2}
\end{multline}
where $\zeta'(-1,\chi)$ is the first derivative of the Hurwitz zeta function~\eqref{eq:zetas-1}.
Using Eqs.~\eqref{eq:I1} and \eqref{eq:I2} in \Eqref{eq:Int1}, we finally obtain
\begin{multline}
 \int_{0}^\infty{\rm d}\tau\, K_0(\tau)\biggl[\ln(\tfrac{\tau}{\tilde\eta})+\gamma-\frac{5}{6}\biggr] \\
=-\frac{1}{3}\Bigl[(1-\kappa^2)\ln\tilde\eta+\kappa^2\tilde\eta\Bigr]
  + {\cal O}\bigl((\tfrac{1}{\tilde\eta})^0\bigr){\cal O}(\kappa^0) + {\cal O}(\kappa^4)\,. \label{eq:Int1_v2}
\end{multline}
Putting everything together, we hence have
\begin{equation}
 \frac{{\cal L}_\text{HE}^{2\text{-loop}}\big|_{1{\rm PI}}}{m^4}=\frac{\alpha}{\pi}\frac{\tilde\eta^2}{32\pi^2}\biggl\{(1-\kappa^2)\ln\tilde\eta
 -\frac{2}{3}\kappa^2\tilde\eta
  + {\cal O}\bigl((\tfrac{1}{\tilde\eta})^0\bigr){\cal O}(\kappa^0) + {\cal O}(\kappa^4)\biggr\} \,. \label{eq:L2sf}
\end{equation}
In a last step, we employ [cf. \Eqref{eq:lims}]
\begin{equation}
 \kappa^2=\frac{\tilde{\cal G}^2}{(2\tilde{\cal F})^2}+{\cal O}\bigl((\tfrac{\tilde{\cal G}^2}{\tilde{\cal F}^2})^2\bigr) \quad\text{and}\quad \tilde\eta=\sqrt{2\tilde{\cal F}}\Bigl(1+{\cal O}\bigl((\tfrac{\tilde{\cal G}^2}{\tilde{\cal F}^2})^1\bigr)\Bigr)\,
\end{equation}
to write \Eqref{eq:L2sf} in the form of \Eqref{eq:L2loopstrongfield} in the main text.


\begin{thebibliography}{10}\setlength{\itemsep}{-0.5mm}

%\cite{Euler:1935zz}
\bibitem{Euler:1935zz} 
  H.~Euler and B.~Kockel,
  %``\"Uber die Streuung von Licht an Licht nach der Diracschen Theorie,''
  Naturwiss.\  {\bf 23}, 246 (1935).
  %%CITATION = NATWA,23,246;%%
  %137 citations counted in INSPIRE as of 23 Feb 2015

%\cite{Heisenberg:1935qt}
\bibitem{Heisenberg:1935qt} 
  W.~Heisenberg and H.~Euler,
  %``Folgerungen aus der Diracschen Theorie des Positrons,''
  Z.\ Phys.\  {\bf 98}, 714 (1936), 
  an English translation is available at [physics/0605038].
  %%CITATION = PHYSICS/0605038;%%
  %%CITATION = ZEPYA,98,714;%%
  %879 citations counted in INSPIRE as of 06 May 2013

%\cite{Weisskopf}
\bibitem{Weisskopf}
V.~Weisskopf,
%``\"Uber die Elektrodynamik des Vakuums auf Grund der Quanthentheorie des Elektrons,''
Kong.\ Dans.\ Vid.\ Selsk., Mat.-fys.\ Medd.\ {\bf XIV}, 6 (1936).

%\cite{Dittrich:1985yb}
\bibitem{Dittrich:1985yb} 
  W.~Dittrich and M.~Reuter,
  %``Effective Lagrangians In Quantum Electrodynamics,''
  Lect.\ Notes Phys.\  {\bf 220}, 1 (1985).
  %%CITATION = LNPHA,220,1;%%
  %56 citations counted in INSPIRE as of 21 août 2015

%\cite{Dittrich:2000zu}
\bibitem{Dittrich:2000zu} 
  W.~Dittrich and H.~Gies,
  %``Probing the quantum vacuum. Perturbative effective action approach in quantum electrodynamics and its application,''
  Springer Tracts Mod.\ Phys.\  {\bf 166}, 1 (2000).
  %doi:10.1007/3-540-45585-X
  %%CITATION = doi:10.1007/3-540-45585-X;%%
  %131 citations counted in INSPIRE as of 24 Nov 2015

%\cite{Marklund:2008gj}
\bibitem{Marklund:2008gj} 
  M.~Marklund and J.~Lundin,
  %``Quantum Vacuum Experiments Using High Intensity Lasers,''
  Eur.\ Phys.\ J.\ D {\bf 55}, 319 (2009)
  [arXiv:0812.3087 [hep-th]].
  %%CITATION = ARXIV:0812.3087;%%
  %18 citations counted in INSPIRE as of 06 May 2013

%\cite{Dunne:2008kc}
\bibitem{Dunne:2008kc} 
  G.~V.~Dunne,
  %``New Strong-Field QED Effects at ELI: Nonperturbative Vacuum Pair Production,''
  Eur.\ Phys.\ J.\ D {\bf 55}, 327 (2009)
  [arXiv:0812.3163 [hep-th]].
  %%CITATION = ARXIV:0812.3163;%%
  %45 citations counted in INSPIRE as of 06 May 2013

%\cite{Heinzl:2008an}
\bibitem{Heinzl:2008an} 
  T.~Heinzl and A.~Ilderton,
  %``Exploring high-intensity QED at ELI,''
  Eur.\ Phys.\ J.\ D {\bf 55}, 359 (2009)
  [arXiv:0811.1960 [hep-ph]].
  
%\cite{DiPiazza:2011tq}
\bibitem{DiPiazza:2011tq} 
  A.~Di Piazza, C.~Muller, K.~Z.~Hatsagortsyan and C.~H.~Keitel,
  %``Extremely high-intensity laser interactions with fundamental quantum systems,''
  Rev.\ Mod.\ Phys.\  {\bf 84}, 1177 (2012)
  [arXiv:1111.3886 [hep-ph]].
  %%CITATION = ARXIV:1111.3886;%%
  %34 citations counted in INSPIRE as of 06 May 2013

%\cite{Dunne:2012vv}
\bibitem{Dunne:2012vv} 
  G.~V.~Dunne,
  %``The Heisenberg-Euler Effective Action: 75 years on,''
  Int.\ J.\ Mod.\ Phys.\ A {\bf 27}, 1260004 (2012)
  [Int.\ J.\ Mod.\ Phys.\ Conf.\ Ser.\  {\bf 14}, 42 (2012)]
  [arXiv:1202.1557 [hep-th]].
  %%CITATION = ARXIV:1202.1557;%%
  %9 citations counted in INSPIRE as of 12 Jan 2015
  
%\cite{Battesti:2012hf}
\bibitem{Battesti:2012hf} 
  R.~Battesti and C.~Rizzo,
  %``Magnetic and electric properties of quantum vacuum,''
  Rept.\ Prog.\ Phys.\  {\bf 76}, 016401 (2013)
  [arXiv:1211.1933 [physics.optics]].
  %%CITATION = ARXIV:1211.1933;%%
  %13 citations counted in INSPIRE as of 01 Dec 2014
  
%\cite{King:2015tba}
\bibitem{King:2015tba} 
  B.~King and T.~Heinzl,
  %``Measuring Vacuum Polarisation with High Power Lasers,''
  High Power Laser Science and Engineering, 4, e5 (2016)
  %doi:10.1017/hpl.2016.1
  [arXiv:1510.08456 [hep-ph]].
  %%CITATION = doi:10.1017/hpl.2016.1;%%
  %8 citations counted in INSPIRE as of 21 Nov 2016
 
%\cite{Karbstein:2016hlj}
\bibitem{Karbstein:2016hlj} 
  F.~Karbstein,
  %``The quantum vacuum in electromagnetic fields: From the Heisenberg-Euler effective action to vacuum birefringence,''
  %doi:10.3204/DESY-PROC-2016-04
  arXiv:1611.09883 [hep-th].
  %%CITATION = doi:10.3204/DESY-PROC-2016-04;%%

%\cite{Sauter:1931zz}
\bibitem{Sauter:1931zz} 
  F.~Sauter,
  %``Uber das Verhalten eines Elektrons im homogenen elektrischen Feld nach der relativistischen Theorie Diracs,''
  Z.\ Phys.\ {\bf 69}, 742 (1931).
  %doi:10.1007/BF01339461
  %%CITATION = doi:10.1007/BF01339461;%%
  %385 citations counted in INSPIRE as of 05 Aug 2016

%\cite{Schwinger:1951nm}
\bibitem{Schwinger:1951nm} 
  J.~S.~Schwinger,
  %``On gauge invariance and vacuum polarization,''
  Phys.\ Rev.\ {\bf 82}, 664 (1951).
  %%CITATION = PHRVA,82,664;%%
  %3762 citations counted in INSPIRE as of 09 Oct 2015
 
  
%\cite{Toll:1952}
\bibitem{Toll:1952}
J.~S.~Toll,
%``The dispersion relation for light and its application to problems involving electron pairs,''
Ph.D. thesis, Princeton Univ., 1952 (unpublished).

%\cite{Baier}
\bibitem{Baier}
R.~Baier and P.~Breitenlohner,
%``Photon Propagation in external fields,''
{Act.~Phys.~Austriaca} {\bf 25}, 212 (1967); 
%R.~Baier and P.~Breitenlohner,
%``The vacuum refraction index in the presence of external fields,''
{Nuov.~Cim.~B}\ {\bf 47} 117 (1967).

%\cite{BialynickaBirula:1970vy}
\bibitem{BialynickaBirula:1970vy} 
  Z.~Bialynicka-Birula and I.~Bialynicki-Birula,
  %``Nonlinear effects in Quantum Electrodynamics. Photon propagation and photon splitting in an external field,''
  Phys.\ Rev.\ D {\bf 2}, 2341 (1970).
  %%CITATION = PHRVA,D2,2341;%%
  %164 citations counted in INSPIRE as of 16 Dec 2014

%\cite{Cantatore:2008zz}
\bibitem{Cantatore:2008zz} 
  G.~Cantatore [PVLAS Collaboration],
  %``Recent results from the PVLAS experiment on the magnetized vacuum,''
  Lect.\ Notes Phys.\  {\bf 741}, 157 (2008); %.
  %%CITATION = LNPHA,741,157;%%
  %3 citations counted in INSPIRE as of 06 May 2013
%\cite{Zavattini:2007ee}
%\bibitem{Zavattini:2007ee} 
  E.~Zavattini {\it et al.} [PVLAS Collaboration],
  %``New PVLAS results and limits on magnetically induced optical rotation and ellipticity in vacuum,''
  Phys.\ Rev.\ D {\bf 77}, 032006 (2008);
  %[arXiv:0706.3419 [hep-ex]]; %.
  %%CITATION = ARXIV:0706.3419;%%
  %130 citations counted in INSPIRE as of 06 May 2013
%\cite{DellaValle:2013xs}
%\bibitem{DellaValle:2013xs} 
  F.~Della Valle {\it et al.},
  %``Measurements of vacuum magnetic birefringence using permanent dipole magnets: the PVLAS experiment,''
  New\ J.\ Phys.\ {\bf 15} 053026 (2013).
  %arXiv:1301.4918 [quant-ph].
  %%CITATION = ARXIV:1301.4918;%%

%\cite{Berceau:2011zz}
\bibitem{Berceau:2011zz} 
  P.~Berceau, R.~Battesti, M.~Fouche and C.~Rizzo,
  %``The vacuum magnetic birefringence experiment: A test for quantum electrodynamics,''
  Can.\ J.\ Phys.\  {\bf 89}, 153 (2011);
  %%CITATION = CJPHA,89,153;%%
  P.~Berceau, M.~Fouche, R.~Battesti and C.~Rizzo,
  %``Magnetic linear birefringence measurements using pulsed fields,''
  Phys.\ Rev.\ A, {\bf 85}, 013837 (2012);
  %[arXiv:1109.4792 [physics.optics]];
  A.~Cadene, P.~Berceau, M.~Fouche, R.~Battesti and C.~Rizzo,
  %``Vacuum magnetic linear birefringence using pulsed fields: status of the BMV experiment,''
  Eur.\ Phys.\ J.\ D {\bf 68}, 16 (2014).
  %[arXiv:1302.5389 [physics.optics]].
  %%CITATION = ARXIV:1302.5389;%%

%\cite{Kotkin:1996nf}
\bibitem{Kotkin:1996nf} 
  G.~L.~Kotkin and V.~G.~Serbo,
  %``Variation in polarization of high-energy gamma quanta traversing a bunch of polarized laser photons,''
  Phys.\ Lett.\ B {\bf 413}, 122 (1997).
  %[hep-ph/9611345].
  %%CITATION = HEP-PH/9611345;%%
  %24 citations counted in INSPIRE as of 29 Jun 2015
  
%\cite{Heinzl:2006xc}
\bibitem{Heinzl:2006xc} 
  T.~Heinzl, B.~Liesfeld, K.~U.~Amthor, H.~Schwoerer, R.~Sauerbrey and A.~Wipf,
  %``On the observation of vacuum birefringence,''
  Opt.\ Commun.\  {\bf 267}, 318 (2006).
  %[hep-ph/0601076].
  %%CITATION = HEP-PH/0601076;%%
  %92 citations counted in INSPIRE as of 16 Apr 2015
  
%\cite{DiPiazza:2006pr}
\bibitem{DiPiazza:2006pr} 
  A.~Di Piazza, K.~Z.~Hatsagortsyan and C.~H.~Keitel,
  %``Light diffraction by a strong standing electromagnetic wave,''
  Phys.\ Rev.\ Lett.\  {\bf 97}, 083603 (2006)
  [hep-ph/0602039].
  %%CITATION = HEP-PH/0602039;%%
  %53 citations counted in INSPIRE as of 01 Dec 2014

%\cite{Dinu:2013gaa}
\bibitem{Dinu:2013gaa} 
  V.~Dinu, T.~Heinzl, A.~Ilderton, M.~Marklund and G.~Torgrimsson,
  %``Vacuum refractive indices and helicity flip in strong-field QED,''
  Phys.\ Rev.\ D {\bf 89}, 125003 (2014)
  [arXiv:1312.6419 [hep-ph]]; %.
  %%CITATION = ARXIV:1312.6419;%%
  %7 citations counted in INSPIRE as of 01 Dec 2014
%\cite{Dinu:2014tsa}
%\bibitem{Dinu:2014tsa} 
%  V.~Dinu, T.~Heinzl, A.~Ilderton, M.~Marklund and G.~Torgrimsson,
  %``Photon polarisation in light-by-light scattering: finite size effects,''
  Phys.\ Rev.\ D {\bf 90}, 045025 (2014)
  [arXiv:1405.7291 [hep-ph]].
  %%CITATION = ARXIV:1405.7291;%%
  %6 citations counted in INSPIRE as of 01 Dec 2014
  
%\cite{Ilderton:2016khs}
\bibitem{Ilderton:2016khs} 
  A.~Ilderton and M.~Marklund,
  %``Prospects for studying vacuum polarisation using dipole and synchrotron radiation,''
  J.\ Plasma Phys.\  {\bf 82}, 655820201 (2016)
  %doi:10.1017/S0022377816000192
  [arXiv:1601.08045 [hep-ph]].
  %%CITATION = doi:10.1017/S0022377816000192;%%
  %1 citations counted in INSPIRE as of 25 May 2016

%\cite{King:2016jnl}
\bibitem{King:2016jnl} 
  B.~King and N.~Elkina,
  %``Vacuum birefringence in high-energy laser-electron collisions,''
  arXiv:1603.06946 [hep-ph].
  %%CITATION = ARXIV:1603.06946;%%

%\cite{Schlenvoigt:2016}
\bibitem{Schlenvoigt:2016} 
 H.~-P.~Schlenvoigt, T.~Heinzl, U.~Schramm, T.~Cowan and R.~Sauerbrey,
 %``Detecting vacuum birefringence with x-ray free electron lasers and high-power optical lasers: a feasibility study,''
 Physica\ Scripta {\bf 91}, 023010 (2016).
  
%\cite{Karbstein:2015xra}
\bibitem{Karbstein:2015xra} 
  F.~Karbstein, H.~Gies, M.~Reuter and M.~Zepf,
  %``Vacuum birefringence in strong inhomogeneous electromagnetic fields,''
  Phys.\ Rev.\ D {\bf 92}, 071301 (2015)
  [arXiv:1507.01084 [hep-ph]].
  %%CITATION = ARXIV:1507.01084;%%
  %1 citations counted in INSPIRE as of 02 Nov 2015
  
%\cite{Karbstein:2016lby}
\bibitem{Karbstein:2016lby} 
  F.~Karbstein and C.~Sundqvist,
  %``Probing vacuum birefringence using x-ray free electron and optical high-intensity lasers,''
  Phys.\ Rev.\ D {\bf 94}, 013004 (2016)
  %doi:10.1103/PhysRevD.94.013004
  [arXiv:1605.09294 [hep-ph]].
  %%CITATION = doi:10.1103/PhysRevD.94.013004;%%
  %1 citations counted in INSPIRE as of 21 Nov 2016

%\cite{Mignani:2016fwz}
\bibitem{Mignani:2016fwz} 
  R.~P.~Mignani, V.~Testa, D.~G.~Caniulef, R.~Taverna, R.~Turolla, S.~Zane and K.~Wu,
  %``Evidence for vacuum birefringence from the first optical polarimetry measurement of the isolated neutron star RX\, J1856.5$-$3754,''
  Mon.\ Not.\ R.\ Astron.\ Soc. {\bf 465} 492-500 (2017)
  %doi:10.1093/mnras/stw2798
  arXiv:1610.08323 [astro-ph.HE].
  %%CITATION = doi:10.1093/mnras/stw2798;%%
 
%\cite{Lundstrom:2005za}
\bibitem{Lundstrom:2005za} 
  E.~Lundstrom, G.~Brodin, J.~Lundin, M.~Marklund, R.~Bingham, J.~Collier, J.~T.~Mendonca and P.~Norreys,
  %``Using high-power lasers for detection of elastic photon-photon scattering,''
  Phys.\ Rev.\ Lett.\  {\bf 96}, 083602 (2006)
  %doi:10.1103/PhysRevLett.96.083602
  [hep-ph/0510076].
  %%CITATION = doi:10.1103/PhysRevLett.96.083602;%%
  %54 citations counted in INSPIRE as of 14 Jan 2016
 
%\cite{Lundin:2006wu}
\bibitem{Lundin:2006wu} 
  J.~Lundin, M.~Marklund, E.~Lundstrom, G.~Brodin, J.~Collier, R.~Bingham, J.~T.~Mendonca and P.~Norreys,
  %``Detection of elastic photon-photon scattering through four-wave mixing using high power lasewie?rs,''
  Phys.\ Rev.\ A {\bf 74}, 043821 (2006)
  [hep-ph/0606136].
  %%CITATION = HEP-PH/0606136;%%
  %18 citations counted in INSPIRE as of 16 mar 2015
  
%\cite{Tommasini:2010fb} [33] FELIX
\bibitem{Tommasini:2010fb} 
  D.~Tommasini and H.~Michinel,
  %``Light by light diffraction in vacuum,''
  Phys.\ Rev.\ A {\bf 82}, 011803 (2010)
  [arXiv:1003.5932 [hep-ph]].
  %%CITATION = ARXIV:1003.5932;%%
  %6 citations counted in INSPIRE as of 06 May 2013

%\cite{King:2012aw}
\bibitem{King:2012aw} 
  B.~King and C.~H.~Keitel,
  %``Photon-photon scattering in collisions of laser pulses,''
  New J.\ Phys.\  {\bf 14}, 103002 (2012)
  [arXiv:1202.3339 [hep-ph]].
  %%CITATION = ARXIV:1202.3339;%%
  %9 citations counted in INSPIRE as of 08 Nov 2014

%\cite{King:2013am}
\bibitem{King:2013am} 
  B.~King, A.~Di Piazza and C.~H.~Keitel,
  %``A matterless double slit,''
 Nature Photon.\ {\bf 4}, 92 (2010) 
  [arXiv:1301.7038 [physics.optics]]; %.
  %%CITATION = ARXIV:1301.7038;%%
%\cite{King:2013zz}
%\bibitem{King:2013zz} 
%  B.~King, A.~Di Piazza and C.~H.~Keitel,
  %``Double-slit vacuum polarisation effects in ultra-intense laser fields,''
Phys.\ Rev.\ A {\bf 82}, 032114 (2010)
  [arXiv:1301.7008 [physics.optics]].
  %%CITATION = ARXIV:1301.7008;%%

%\cite{Hatsagortsyan:2011}
\bibitem{Hatsagortsyan:2011}
K.~Z.~Hatsagortsyan and G.~Y.~Kryuchkyan,
%``Bragg Scattering of Light in Vacuum Structured by Strong Periodic Fields,''
 Phys.\ Rev.\ Lett. {\bf 107}, 053604 (2011).
  
%\cite{Sarazin:2016zer}
\bibitem{Sarazin:2016zer} 
  X.~Sarazin, F.~Couchot, A.~Djannati-Atai, O.~Guilbaud, S.~Kazamias, M.~Pittman and M.~Urban,
  %``Refraction of light by light in vacuum,''
  Eur.\ Phys.\ J.\ D {\bf 70}, 13 (2016).
  %doi:10.1140/epjd/e2015-60428-5
  %%CITATION = doi:10.1140/epjd/e2015-60428-5;%%
  %2 citations counted in INSPIRE as of 05 Aug 2016 

%\cite{Gies:2013yxa}
\bibitem{Gies:2013yxa} 
  H.~Gies, F.~Karbstein and N.~Seegert,
  %``Quantum Reflection as a New Signature of Quantum Vacuum Nonlinearity,''
  New J.\ Phys.\  {\bf 15}, 083002 (2013)
  [arXiv:1305.2320 [hep-ph]];
  %%CITATION = ARXIV:1305.2320;%%
  %9 citations counted in INSPIRE as of 10 Dec 2014jia
%\cite{Gies:2014wsa}
%\bibitem{Gies:2014wsa} 
%  H.~Gies, F.~Karbstein and N.~Seegert,
  %``Quantum reflection of photons off spatio-temporal electromagnetic field inhomogeneities,''
  New J.\ Phys.\  {\bf 17}, 043060 (2015)
  [arXiv:1412.0951 [hep-ph]].
  %%CITATION = doi:10.1088/1367-2630/17/4/043060;%%
  %4 citations counted in INSPIRE as of 02 Dec 2015  

%\cite{Yakovlev:1935qt}
\bibitem{Yakovlev:1966} 
  V.P.~Yakovlev,
  %``Incoherent electromagnetic wave scattering in a Coulomb field,''
  Sov. \ Phys. \ JETP \ {\bf 24}, 411 (1967)\ [Zh.\ Eksp.\ Teor. \ Fiz. {\bf 51}, 619 (1966)]. 
  
%\cite{DiPiazza:2007cu}
\bibitem{DiPiazza:2007cu} 
  A.~Di Piazza, K.~Z.~Hatsagortsyan and C.~H.~Keitel,
  %``Non-perturbative vacuum-polarization effects in proton-laser collisions,''
  Phys.\ Rev.\ Lett.\  {\bf 100}, 010403 (2008)
  %doi:10.1103/PhysRevLett.100.010403
  [arXiv:0708.0475 [hep-ph]];
  %%CITATION = doi:10.1103/PhysRevLett.100.010403;%%
  %20 citations counted in INSPIRE as of 02 Dec 2015
%\cite{DiPiazza:2009cq}
%\bibitem{DiPiazza:2009cq} 
%  A.~Di Piazza, K.~Z.~Hatsagortsyan and C.~H.~Keitel,
  %``Laser photon merging in proton-laser collisions,''
  Phys.\ Rev.\ A {\bf 78}, 062109 (2008)
  %doi:10.1103/PhysRevA.78.062109
  [arXiv:0906.5576 [hep-ph]].
  %%CITATION = doi:10.1103/PhysRevA.78.062109;%%
  %6 citations counted in INSPIRE as of 02 Dec 2015
 
%\cite{Gies:2014jia}
\bibitem{Gies:2014jia} 
  H.~Gies, F.~Karbstein and R.~Shaisultanov,
  %``Laser photon merging in an electromagnetic field inhomogeneity,''
  Phys.\ Rev.\ D {\bf 90}, 033007 (2014)
  [arXiv:1406.2972 [hep-ph]].
  %%CITATION = ARXIV:1406.2972;%%
  %3 citations counted in INSPIRE as of 08 Dec 2014

%\cite{Gies:2016czm}
\bibitem{Gies:2016czm} 
  H.~Gies, F.~Karbstein and N.~Seegert,
  %``Photon merging and splitting in electromagnetic field inhomogeneities,''
  Phys.\ Rev.\ D {\bf 93}, 085034 (2016)
  %doi:10.1103/PhysRevD.93.085034
  [arXiv:1603.00314 [hep-ph]].
  %%CITATION = doi:10.1103/PhysRevD.93.085034;%%
  %1 citations counted in INSPIRE as of 07 May 2016  
  
%\cite{Adler:1971wn}
\bibitem{Adler:1971wn}
  S.~L.~Adler,
  %``Photon Splitting And Photon Dispersion In A Strong Magnetic Field,''
  Annals\ Phys.\  {\bf 67}, 599 (1971).
  %%CITATION = APNYA,67,599;%%
  
%\cite{Adler:1970gg}
\bibitem{Adler:1970gg} 
  S.~L.~Adler, J.~N.~Bahcall, C.~G.~Callan and M.~N.~Rosenbluth,
  %``Photon splitting in a strong magnetic field,''
  Phys.\ Rev.\ Lett.\  {\bf 25}, 1061 (1970).
  %doi:10.1103/PhysRevLett.25.1061
  %%CITATION = doi:10.1103/PhysRevLett.25.1061;%%
  %82 citations counted in INSPIRE as of 02 Dec 2015
 
%\cite{Papanyan:1971cv}
\bibitem{Papanyan:1971cv} 
  V.~O.~Papanyan and V.~I.~Ritus,
  %``Vacuum polarization and photon splitting in an intense field,''
  Sov. \ Phys. \ JETP \ {\bf 34}, 1195 (1972) \ [Zh.\ Eksp.\ Teor.\ Fiz.\  {\bf 61}, 2231 (1971)];
  %%CITATION = ZETFA,61,2231;%%
  %8 citations counted in INSPIRE as of 26 Mar 2015
%\cite{Papanyan:1973xa}
%\bibitem{Papanyan:1973xa} 
%  V.~O.~Papanyan and V.~I.~Ritus,
  %``Three-photon interaction in an intense field and scaling invariance,''
  Sov. \ Phys. \ JETP \ {\bf 38}, 879 (1974) \ [Zh.\ Eksp.\ Teor.\ Fiz.\  {\bf 65}, 1756 (1973)].
  %%CITATION = ZETFA,65,1756;%%
  %4 citations counted in INSPIRE as of 10 Dec 2014
  
%\cite{Stoneham:1979}
\bibitem{Stoneham:1979}
  R.~J.~Stoneham,
  %``Phonon splitting in the magnetised vacuum,''
  J.\ Phys. A, {\bf 12}, 2187 (1979).
%  url={http://stacks.iop.org/0305-4470/12/i=11/a=028},

%\cite{Baier:1986cv}
\bibitem{Baier:1986cv} 
  V.~N.~Baier, A.~I.~Milshtein and R.~Z.~Shaisultanov,
  %``Photon Splitting in a Strong Electromagnetic Field,''
  Sov.\ Phys.\ JETP {\bf 63}, 665 (1986) \  [Zh.\ Eksp.\ Teor.\ Fiz.\  {\bf 90}, 1141 (1986)];
  %%CITATION = SPHJA,63,665;%%
  %10 citations counted in INSPIRE as of 19 Oct 2015
%\cite{Baier:1996bq}
%\bibitem{Baier:1996bq} 
%  V.~N.~Baier, A.~I.~Milshtein and R.~Z.~Shaisultanov,
  %``Photon splitting in a very strong magnetic field,''
  Phys.\ Rev.\ Lett.\  {\bf 77}, 1691 (1996)
  %doi:10.1103/PhysRevLett.77.1691
  [hep-th/9604028].
  %%CITATION = doi:10.1103/PhysRevLett.77.1691;%%
  %43 citations counted in INSPIRE as of 02 Dec 2015

%\cite{Adler:1996cja}
\bibitem{Adler:1996cja} 
  S.~L.~Adler and C.~Schubert,
  %``Photon splitting in a strong magnetic field: Recalculation and comparison with previous calculations,''
  Phys.\ Rev.\ Lett.\  {\bf 77}, 1695 (1996)
  %doi:10.1103/PhysRevLett.77.1695
  [hep-th/9605035].
  %%CITATION = doi:10.1103/PhysRevLett.77.1695;%%
  %77 citations counted in INSPIRE as of 02 Dec 2015 

%\cite{DiPiazza:2007yx}
\bibitem{DiPiazza:2007yx} 
  A.~Di Piazza, A.~I.~Milstein and C.~H.~Keitel,
  %``Photon splitting in a laser field,''
  Phys.\ Rev.\ A {\bf 76}, 032103 (2007)
  [arXiv:0704.0695 [hep-ph]].
  %%CITATION = ARXIV:0704.0695;%%
  %17 citations counted in INSPIRE as of 19 Oct 2015

%\cite{DiPiazza:2005jc}
\bibitem{DiPiazza:2005jc} 
  A.~Di Piazza, K.~Z.~Hatsagortsyan and C.~H.~Keitel,
  %``Harmonic generation from laser-driven vacuum,''
  Phys.\ Rev.\ D {\bf 72}, 085005 (2005).
  %doi:10.1103/PhysRevD.72.085005
  %%CITATION = doi:10.1103/PhysRevD.72.085005;%%
  %30 citations counted in INSPIRE as of 05 Aug 2016

%\cite{Fedotov:2006ii} [51] LENA
\bibitem{Fedotov:2006ii} 
  A.~M.~Fedotov and N.~B.~Narozhny,
  %``Generation of harmonics by a focused laser beam in vacuum,''
  Phys.\ Lett.\ A {\bf 362}, 1 (2007)
  %doi:10.1016/j.physleta.2006.09.085
  [hep-ph/0604258].
  %%CITATION = doi:10.1016/j.physleta.2006.09.085;%%
  %17 citations counted in INSPIRE as of 05 Aug 2016

%\cite{Karbstein:2014fva}
\bibitem{Karbstein:2014fva} 
  F.~Karbstein and R.~Shaisultanov,
  %``Stimulated photon emission from the vacuum,''
  Phys.\ Rev.\ D {\bf 91}, 113002 (2015)
  [arXiv:1412.6050 [hep-ph]].
  %%CITATION = ARXIV:1412.6050;%%
  %1 citations counted in INSPIRE as of 16 Apr 2015
  
%\cite{Bohl:2015uba}
\bibitem{Bohl:2015uba} 
  P.~B\"ohl, B.~King and H.~Ruhl,
  %``Vacuum high harmonic generation in the shock regime,''
  Phys.\ Rev.\ A {\bf 92}, 032115 (2015)
  %doi:10.1103/PhysRevA.92.032115
  [arXiv:1503.05192 [physics.plasm-ph]].
  %%CITATION = doi:10.1103/PhysRevA.92.032115;%%
  %4 citations counted in INSPIRE as of 05 Aug 2016
  
\bibitem{Ritus:1975} 
  V.~I.~Ritus,
  %``Lagrangian of an intense electromagnetic field and quantum electrodynamics at short distances,''
  Sov.\ Phys.\ JETP\ {\bf 42}, 774 (1975) \ [Zh.\ Eksp.\ Teor.\ Fiz.\  {\bf 69}, 1517 (1975)].

%\cite{Dittrich:1998fy}
\bibitem{Dittrich:1998fy} 
  W.~Dittrich and H.~Gies,
  %``Light propagation in nontrivial QED vacua,''
  Phys.\ Rev.\ D {\bf 58}, 025004 (1998)
  %doi:10.1103/PhysRevD.58.025004
  [hep-ph/9804375].
  %%CITATION = doi:10.1103/PhysRevD.58.025004;%%
  %107 citations counted in INSPIRE as of 19 Dec 2016
  
\bibitem{Ritus:1977} 
  V.~I.~Ritus,
  %``Connection between strong-field quantum electrodynamics with shortdistance quantum electrodynamics,''
  Sov.\ Phys.\ JETP\ {\bf 46}, 423 (1977) \ [Zh.\ Eksp.\ Teor.\ Fiz.\  {\bf 73}, 807 (1977)].

%\cite{Karbstein:2015cpa}
\bibitem{Karbstein:2015cpa} 
  F.~Karbstein and R.~Shaisultanov,
  %``Photon propagation in slowly varying inhomogeneous electromagnetic fields,''
  Phys.\ Rev.\ D {\bf 91}, 085027 (2015)
  [arXiv:1503.00532 [hep-ph]].
  %%CITATION = ARXIV:1503.00532;%%
  %1 citations counted in INSPIRE as of 24 Apr 2015
  
%\cite{Fliegner:1997ra}
\bibitem{Fliegner:1997ra} 
  D.~Fliegner, M.~Reuter, M.~G.~Schmidt and C.~Schubert,
  %``The Two loop Euler-Heisenberg Lagrangian in dimensional renormalization,''
  Theor.\ Math.\ Phys.\  {\bf 113}, 1442 (1997)
  [Teor.\ Mat.\ Fiz.\  {\bf 113}, 289 (1997)]
  %doi:10.1007/BF02634170
  [hep-th/9704194].
  %%CITATION = doi:10.1007/BF02634170;%%
  %53 citations counted in INSPIRE as of 18 Dec 2016

%\cite{Kors:1998ew}
\bibitem{Kors:1998ew} 
  B.~Kors and M.~G.~Schmidt,
  %``The Effective two loop Euler-Heisenberg action for scalar and spinor QED in a general constant background field,''
  Eur.\ Phys.\ J.\ C {\bf 6}, 175 (1999)
  %doi:10.1007/s100500050332, 10.1007/s100520050331
  [hep-th/9803144].
  %%CITATION = doi:10.1007/s100500050332, 10.1007/s100520050331;%%
  %43 citations counted in INSPIRE as of 18 Dec 2016
  
%\cite{Dunne:2004nc}
\bibitem{Dunne:2004nc} 
  G.~V.~Dunne,
  %``Heisenberg-Euler effective Lagrangians: Basics and extensions,''
  In *Shifman, M. (ed.) et al.: From fields to strings, vol. 1* 445-522
  [hep-th/0406216].
  %%CITATION = HEP-TH/0406216;%%
  %187 citations counted in INSPIRE as of 19 août 2015

%\cite{Dunne:2002ta}
\bibitem{Dunne:2002ta} 
  G.~V.~Dunne, H.~Gies and C.~Schubert,
  %``Zero modes, beta functions and IR / UV interplay in higher loop QED,''
  JHEP {\bf 0211}, 032 (2002)
  %doi:10.1088/1126-6708/2002/11/032
  [hep-th/0210240].
  %%CITATION = doi:10.1088/1126-6708/2002/11/032;%%
  %25 citations counted in INSPIRE as of 19 Dec 2016

%\cite{Cangemi:1995ee}
\bibitem{Cangemi:1995ee} 
  D.~Cangemi, E.~D'Hoker and G.~V.~Dunne,
  %``Effective energy for QED in (2+1)-dimensions with semilocalized magnetic fields: A Solvable model,''
  Phys.\ Rev.\ D {\bf 52}, 3163 (1995)
  [hep-th/9506085].
  %%CITATION = HEP-TH/9506085;%%
  %36 citations counted in INSPIRE as of 09 Oct 2015
  
%\cite{Dunne:1997kw}
\bibitem{Dunne:1997kw} 
  G.~V.~Dunne and T.~M.~Hall,
  %``An exact (3+1)-dimensional QED effective action,''
  Phys.\ Lett.\ B {\bf 419}, 322 (1998)
  [hep-th/9710062].
  %%CITATION = HEP-TH/9710062;%%
  %36 citations counted in INSPIRE as of 09 Oct 2015

%\cite{Dunne:1998ni}
\bibitem{Dunne:1998ni} 
  G.~V.~Dunne and T.~Hall,
  %``On the QED effective action in time dependent electric backgrounds,''
  Phys.\ Rev.\ D {\bf 58}, 105022 (1998)
  [hep-th/9807031].
  %%CITATION = HEP-TH/9807031;%%
  %63 citations counted in INSPIRE as of 09 Oct 2015  

%\cite{Kim:2009pg}
\bibitem{Kim:2009pg} 
  S.~P.~Kim, H.~K.~Lee and Y.~Yoon,
  %``Effective Action of QED in Electric Field Backgrounds II. Spatially Localized Fields,''
  Phys.\ Rev.\ D {\bf 82}, 025015 (2010)
  %doi:10.1103/PhysRevD.82.025015
  [arXiv:0910.3363 [hep-th]].
  %%CITATION = doi:10.1103/PhysRevD.82.025015;%%
  %31 citations counted in INSPIRE as of 18 Dec 2016

%\cite{Huet:2009cy}
\bibitem{Huet:2009cy} 
  I.~Huet, D.~G.~C.~McKeon and C.~Schubert,
  %``Three-loop Euler-Heisenberg Lagrangian and asymptotic analysis in 1+1 QED,''
  %doi:10.1142/9789814289931\_0064
  arXiv:0911.0227 [hep-th].
  %%CITATION = doi:10.1142/9789814289931_0064;%%
  %2 citations counted in INSPIRE as of 16 Dec 2016

%\cite{Huet:2011kd}
\bibitem{Huet:2011kd} 
  I.~Huet, M.~Rausch de Traubenberg and C.~Schubert,
  %``The Euler-Heisenberg Lagrangian Beyond One Loop,''
  Int.\ J.\ Mod.\ Phys.\ Conf.\ Ser.\  {\bf 14}, 383 (2012)
%  doi:10.1142/S2010194512007507
  [arXiv:1112.1049 [hep-th]].
  %%CITATION = doi:10.1142/S2010194512007507;%%
  %4 citations counted in INSPIRE as of 16 Dec 2016 6
  
%\cite{Galtsov:1982}
\bibitem{Galtsov:1982} 
  G.~V.~Galtsov and N.~S.~Nikitina,
  %``Macroscopic vacuum effects in an inhomogeneous and nonstationary electromagnetic field,''
  Sov.\ Phys.\ JETP\ {\bf 57}, 705 (1983) \ [Zh.\ Eksp.\ Teor.\ Fiz.\  {\bf 84}, 1217 (1983)].

%\cite{Gusynin:1998bt}
\bibitem{Gusynin:1998bt} 
  V.~P.~Gusynin and I.~A.~Shovkovy,
  %``Derivative expansion of the effective action for QED in (2+1)-dimensions and (3+1)-dimensions,''
  J.\ Math.\ Phys.\  {\bf 40}, 5406 (1999)
  %doi:10.1063/1.533037
  [hep-th/9804143].
  %%CITATION = doi:10.1063/1.533037;%%
  %61 citations counted in INSPIRE as of 17 Dec 2016


%\cite{Martin:2003gb}
\bibitem{Martin:2003gb} 
  L.~C.~Martin, C.~Schubert and V.~M.~Villanueva Sandoval,
  %``On the low-energy limit of the QED N photon amplitudes,''
  Nucl.\ Phys.\ B {\bf 668}, 335 (2003)
  [hep-th/0301022].
  %%CITATION = HEP-TH/0301022;%%
  %22 citations counted in INSPIRE as of 20 août 2015
  
%\cite{Galtsov:1971xm}
\bibitem{Galtsov:1971xm} 
  D.~Galtsov and V.~Skobelev,
  %``Photons creation by an external field,''
  Phys.\ Lett.\ B {\bf 36}, 238 (1971).
  %%CITATION = PHLTA,B36,238;%%  
  
%\cite{Karplus:1950zza}
\bibitem{Karplus:1950zz} 
  R.~Karplus and M.~Neuman,
  %``Non-Linear Interactions between Electromagnetic Fields,''
  Phys.\ Rev.\  {\bf 80}, 380 (1950);
  %doi:10.1103/PhysRev.80.380
  %%CITATION = doi:10.1103/PhysRev.80.380;%%
  %152 citations counted in INSPIRE as of 18 Aug 2016
%\cite{Karplus:1950zz}
%\bibitem{Karplus:1950zz} 
  %R.~Karplus and M.~Neuman,
  %``The scattering of light by light,''
  Phys.\ Rev.\  {\bf 83}, 776 (1951).
  %%CITATION = PHRVA,83,776;%%
  %128 citations counted in INSPIRE as of 06 May 2013


\bibitem{Bhartia:1978}
 P.~Bhartia, S.~Valluri,
 %``Non-linear scattering of light in the limit of ultra-strong fields,''
 Can.\ J.\ Phys.\, {\bf 56}, 1122 (1978).

\bibitem{Bhartia:1980}
 S.~R.~Valluri, P.~Bhartia,
 %``An analytical proof for the generation of higher harmonics due to the interaction of plane electromagnetic waves,''
 Can.\ J.\ Phys.\, {\bf 58}, 116 (1980).

%\cite{Gies:2006ca}
\bibitem{Gies:2006ca} 
  H.~Gies, J.~Jaeckel and A.~Ringwald,
  %``Polarized Light Propagating in a Magnetic Field as a Probe of Millicharged Fermions,''
  Phys.\ Rev.\ Lett.\  {\bf 97}, 140402 (2006)
  %doi:10.1103/PhysRevLett.97.140402
  [hep-ph/0607118].
  %%CITATION = doi:10.1103/PhysRevLett.97.140402;%%
  %126 citations counted in INSPIRE as of 09 Aug 2016
 
%\cite{Dittrich:1975au}
\bibitem{Dittrich:1975au} 
  W.~Dittrich,
  %``One Loop Effective Potentials in QED,''
  J.\ Phys.\ A {\bf 9}, 1171 (1976).
  %doi:10.1088/0305-4470/9/7/019
  %%CITATION = doi:10.1088/0305-4470/9/7/019;%%
  %18 citations counted in INSPIRE as of 05 Sep 2016

\bibitem{dlmf:1}
  NIST Digital Library of Mathematical Functions,
  http://dlmf.nist.gov/25.11\#E18, Release 1.0.10 of 2015-08-07.
  
%\cite{Dunne:2001pp}
\bibitem{Dunne:2001pp} 
  G.~V.~Dunne and C.~Schubert,
  %``Closed form two loop Euler-Heisenberg Lagrangian in a selfdual background,''
  Phys.\ Lett.\ B {\bf 526}, 55 (2002)
  %doi:10.1016/S0370-2693(01)01475-7
  [hep-th/0111134].
  %%CITATION = doi:10.1016/S0370-2693(01)01475-7;%%
  %19 citations counted in INSPIRE as of 17 Dec 2016


\bibitem{Gradshteyn}
I.~S.~Gradshteyn and I.~M.~Ryzhik, \textit{Table of Integrals, Series, and Products}, Fifth Edition, Academic Press, UK (1994).

%\cite{Tsai:1975iz}
\bibitem{Tsai:1975iz}
  W.~y.~Tsai and T.~Erber,
  %``The Propagation of Photons in Homogeneous Magnetic Fields: Index of Refraction,''
  Phys.\ Rev.\ D {\bf 12}, 1132 (1975).
  %%CITATION = PHRVA,D12,1132;%%
  %102 citations counted in INSPIRE as of 21 Feb 2015
  
%\cite{Karbstein:2013ufa}
\bibitem{Karbstein:2013ufa} 
  F.~Karbstein,
  %``Photon polarization tensor in a homogeneous magnetic or electric field,''
  Phys.\ Rev.\ D {\bf 88}, 085033 (2013)
  [arXiv:1308.6184 [hep-th]].
  %%CITATION = ARXIV:1308.6184;%%
  %10 citations counted in INSPIRE as of 25 Jan 2015
  
\end{thebibliography}
\end{document}